\documentclass{aa}  

\usepackage{graphicx}
\usepackage{txfonts}
\usepackage{tablefootnote}
\newcommand{\asec}{$^{\prime\prime}$}
\newcommand{\pas}{.\hskip-2pt$^{\prime\prime}$}
\def\SigmaH2{$\Sigma $(${\rm H_2}$)}
\def\r1415{$^{14}$N/$^{15}$N}

\def\H{N$_{2}$H$^{+}$}

\def\15N{$^{15}$NNH$^+$}
\def\N15{N$^{15}$NH$^+$}

\def\H13CN{\mbox{H$^{13}$CN}}

\def\kms{\mbox{km~s$^{-1}$}}

\def\cmq{cm$^{-2}$}

\def\solm{\mbox{M$_\odot$}}

\def\VLSR{$V_{\rm LSR}$}

\def\Ntot{$N_{\rm tot}$}
\def\Tex{\mbox{$T_{\rm ex}$}}

\def\kms{km\,s$^{-1}$}

\begin{document} 

   \title{The GUAPOS project: G31.41+0.31 Unbiased ALMA sPectral Observational Survey.}

   \subtitle{IV. Phosphorus-bearing molecules and their relation with shock tracers}

   \author{F. Fontani
          \inst{1,2,3}
          \and
          C. Mininni\inst{4}
          \and
          M.T. Beltr\'an\inst{1}
          \and
          V.M. Rivilla\inst{5}
          \and
          L. Colzi\inst{5}
          \and
          I. Jim\'enez-Serra\inst{5}
          \and
          \'A. L\'opez-Gallifa\inst{5}
          \and
          \'A. S\'anchez-Monge\inst{6,7}
          \and
          S. Viti\inst{8,9}
          }

   \institute{INAF - Osservatorio Astrofisico di Arcetri,
              Largo E. Fermi 5,
              I-50125, Florence (Italy)\\
              \email{francesco.fontani@inaf.it}
              \and
              Centre for Astrochemical Studies, Max-Planck-Institute for Extraterrestrial Physics, Giessenbachstrasse 1, 85748 Garching, Germany
              \and
              LERMA, Observatoire de Paris, PSL Research University, CNRS, Sorbonne
         Universit\'e, F-92190 Meudon (France)
              \and
             INAF - Istituto di Astrofisica e Planetologia Spaziali, 
             Via Fosso del Cavaliere 100, I-00133 Roma (Italy)
             \and
             Centro de Astrobiolog\'ia (CSIC-INTA), Ctra Ajalvir km 4, 28850, Torrej\'on de Ardoz, Madrid (Spain)
             \and
             Institut de Ci\'encies de l’Espai (ICE, CSIC), Can Magrans s/n, E-08193, Bellaterra, Barcelona, Spain
             \and
             Institut d’Estudis Espacials de Catalunya (IEEC), Barcelona, Spain
             \and
             Leiden Observatory, Leiden University, PO Box 9513, 2300 RA Leiden, The Netherlands
             \and
             Department of Physics and Astronomy, University College London, Gower Street, London, WC1E 6BT, UK
             }

   \date{Received XXX; accepted YYY}

 
  \abstract
   {The astrochemistry of the important biogenic element phosphorus (P) is still poorly understood, but observational evidence indicates that P-bearing molecules are likely associated with shocks.}
   {We study P-bearing molecules, as well as some shock tracers, towards one of the chemically richest hot molecular core, G31.41+0.31, in the framework of the project "G31.41+0.31 Unbiased ALMA sPectral Observational Survey" (GUAPOS), observed with the Atacama Large Millimeter Array (ALMA).}
   {We have observed the molecules PN, PO, SO, SO$_2$, SiO, and SiS, through their rotational lines in the spectral range 84.05--115.91~GHz, covered by the GUAPOS project.}
   {PN is clearly detected while PO is tentatively detected.
   The PN emission arises from two regions southwest of the hot core peak, "1" and "2", and is undetected or tentatively detected towards the hot core peak.
   the PN and SiO lines are very similar both in spatial emission morphology and spectral shape. 
   Region "1" is in part overlapping with the hot core and it is warmer than region "2", which is well separated from the hot core and located along the outflows identified in previous studies.
   The SO, SO$_2$, and SiS emissions are also detected towards the PN emitting regions "1" and "2", but arise mostly from the hot core. Moreover,
the column density ratio SiO/PN remains constant in regions "1" and "2", while SO/PN, SiS/PN, and SO$_2$/PN decrease by about an order of magnitude from region "1" to region "2", indicating that SiO and PN have a common origin even in regions with different physical conditions.
The PO/PN ratio in region "2", where PO is tentatively detected, is $\sim 0.6-0.9$, in line with the predictions of pure shock models.}
   {Our study firmly confirms previous observational evidence that PN emission is tightly associated with SiO and it is likely a product of shock-chemistry, as the lack of a clear detection of PN towards the hot-core allows to rule out relevant formation pathways in hot gas.
   We propose the PN emitting region "2" as a new astrophysical laboratory for shock-chemistry studies.}


   \maketitle
%

\section{Introduction}
\label{intro}

Hot molecular cores (HMCs) are the cradles of future high-mass stars ($M \geq 8$\solm) and rich stellar clusters, and represent the chemically richest environment in the local interstellar medium \citep{Bisschop07,Fontani07,Belloche13,Rivilla17}. 
In fact, many complex astronomical molecules (i.e. molecules with at least 6 atoms), including organic ones (COMs) such as dimethyl ether, ethanol, methyl cyanide, vinyl cyanide, and many others \citep{McGuire18,McGuire22} have been detected for the first time towards the well-known HMCs Orion-KL (Genzel \& Stutzki~\citeyear{GeS89}) and Sgr B2 (e.g.~Bonfand et al.~\citeyear{Bonfand19}).

Such rich chemistry is triggered by several physical processes.
First, HMCs typically harbour one or more deeply embedded protostars which heat up their surrounding molecular gas up to temperatures $T\geq 100$ K. 
In this warm environment, the molecules in the ice mantles of dust grains, formed in the early cold phase, sublimate back in the gas-phase via thermal and
non-thermal desorption (e.g.~Garrod et al.~\citeyear{Garrod22}), and gas reactions not efficient at low temperature start to proceed and form new, more complex molecules \citep{GeH06,Bonfand19,Gieser19,BeG20}. 
Second, collimated jets and molecular outflows from the protostar(s) can trigger local chemistry typical of shocked gas such as grain sputtering and non-equilibrium chemistry
\citep{Hem89,Bachiller96,Tychoniec21}.
Therefore, HMCs with a well-known physical structure are excellent astronomical laboratories to study astrochemistry in a variety of conditions.

In this framework, the project "G31.41+0.31 Unbiased ALMA sPectral Observational Survey (GUAPOS)" \citep{Mininni20} is aimed to observe the full 3mm spectral window accessible with the Atacama Large Millimeter Array (ALMA) towards the well-known HMC G31.41+0.31 (G31 hereafter), located at heliocentric distance of 3.75~kpc (Immer et al.~\citeyear{Immer19}, please see Mininni et al.~\citeyear{Mininni20} for a detailed description of the source).
The unbiased survey allows us to observe many transitions of different species and to confirm (or discard) their detection in a robust way. 
The proper identification of the species and of all their transitions is important also for a proper derivation of physical parameters such as abundances and excitation temperatures, crucial to appropriately constrain the chemical models. 
The project has already provided important constraints on the formation/destruction routes of potential pre-biotic species such as the three isomers of C$_2$H$_4$O$_2$ \citep{Mininni20}, peptide-like bond molecules \citep{Colzi21}, and oxygen- and nitrogen-bearing COMs~\citep{Mininni23,LopezGallifa23}. 
In this work, we focus our analysis on phosphorus-bearing molecules.

Phosphorus (P) is a basic ingredient for life as we know it. P-compounds are unique to form
large biomolecules, thanks to their extreme structural stability and functional reactivity. For these
reasons, it is a crucial component, in the form of phosphates, of nucleic acids (RNA and DNA),
cellular membranes (phospholipids), and the adenosine triphosphate (ATP), the key molecule for
the energy transfers in cells (see e.g. Mac\'ia \citeyear{Macia05}, Pasek et al.~\citeyear{Pasek17}). Despite its low Solar abundance
relative to hydrogen (P/H $\sim 3\times 10^{-7}$, Asplund et al. \citeyear{Asplund09}) with respect to the other biogenic elements such as C, O and N, its abundance 
in living organisms is several orders of magnitude higher than the Solar one, for example up to P/H $\sim 10^{-3}$ in bacteria (e.g. Fagerbakke et
al.~\citeyear{Fagerbakke96}). Therefore, understanding what is the main source and reservoir of P in space, how its
compounds form and evolve, how they are transformed and/or conserved in star-forming regions
and, finally, delivered to planetary systems, is of huge importance in astrophysics. 

P is believed to be mainly formed in massive stars by neutron capture on silicon (Si) in hydrostatic
neon-burning shells in the pre supernova (SN) stage (e.g. Koo et al.~\citeyear{Koo13} and references therein), and
also in explosive carbon- and neon-burning layers during SN explosions (Woosley \& Weaver~\citeyear{WeW95}). 
In the interstellar medium, it is less depleted than previously thought based on observations of P-bearing molecules obtained in massive star-forming regions and evolved stars \citep{Rivilla16,Ziurys18}.
Despite this, a handful of P-bearing molecular species (eight in total, including the tentative detection of SiP) have been detected in the interstellar medium, and only three of them in star-forming regions, namely PO \citep{Rivilla16,Lefloch16,WeB22}, PN \citep{Ziurys87,Fontani16,Rivilla18,Bergner22,WeB22}, and PO$^+$  \citep{Rivilla22}.
Single-dish observations and interferometric maps of PN and PO, the P-bearing molecules which are easier to detect in star-forming regions \citep{Fontani16,Mininni18,Rivilla16,Rivilla18,Bergner22,WeB22}, and evolved stars
\citep{Agundez07,DeBeck13,Ziurys18}, agree with the fact that PN and PO are likely derived from a solid phosphorus carrier, based on their spatial association with SiO emission, a tracer of protostellar outflows and shocks \citep{Lefloch16,Mininni18,Rivilla18,Fontani19, Bernal21}. 

However, the reactions leading to the formation of PN and (especially) PO are far to be clear, even though some theoretical works start to determine the key reactions of the network of phosphorus \citep{Fernandez23}.
For example, \citet{Rivilla20} proposed that to justify the observed abundances, the molecules should not be directly sputtered from the grains but be formed through gas-phase photochemistry induced by ultraviolet (UV) photons from the protostar in post-shocked gas.
\citet{Garcia21} proposed that PO can be formed from atomic P reacting with OH, particularly efficient in warm environments, and \citet{Jimenez18} also proposed mechanisms involving energetic processing.
Moreover, \citet{Rivilla20} and \citet{Bergner22} found that PN and PO emission is co-spatial with low-velocity and not with high-velocity SiO and SO emission, further complicating the simple scenario of pure grain sputtering.

We study here the P-bearing molecules detected in GUAPOS together with other shock tracers: the Silicon-bearing molecules SiO and SiS, and the sulphur-bearing species SO and SO$_2$.
The analysis of the aforementioned sulphur- and silicon-bearing species is limited to the regions in the core where the emission of the phosphorus molecules is significant.
The paper is structured as follows: in Sect.~\ref{obs} we describe the observations and data reduction. In Sect.~\ref{res} we illustrate the observational results, that we discuss in Sect.~\ref{discu}. Conclusions and future perspectives are given in Sect.~\ref{conc}.

\begin{table}[]
    \centering
        \caption{Spectral parameters of the analysed transitions}
        \setlength{\tabcolsep}{0.15cm}
    \begin{tabular}{llccc}
\hline
\hline
Mol.   &   Transition   &  Rest Freq.  & log[$A_{\rm ij}$]  &  $E_{\rm up}$  \\
           &                &   (GHz)            & (s$^{-1}$) &  (K)   \\
\hline
PN & $N=2-1,J=2-2$  &   93.9782  & --5.6144 &  6.8  \\
   & $N=2-1,J=1-0$  &   93.9785  & --5.2677 &  6.8  \\
   & $N=2-1,J=2-1$  &   93.9798  & --5.1373 &  6.8  \\
   & $N=2-1,J=3-2$  &   93.9799  & --5.0123 &  6.8  \\
   & $N=2-1,J=1-2$  &   93.9808  & --6.5687 &  6.8  \\
   & $N=2-1,J=1-1$  &   93.9823  & --5.3926 &  6.8  \\
   &                &            &          &       \\
PO & $J=5/2-3/2,$   &            &          &       \\
   & $\Omega=1/2,$   &          &       \\
   & $\,F=3-2,\,l=e$ & 108.9984 & --4.6712 & 8.4 \\
   & $\,F=2-1,\,l=e$ & 109.0454 & --4.7160 & 8.4 \\
   & $\,F=3-2,\,l=f$ & 109.2062 & --4.6689 & 8.4 \\
   & $\,F=2-1,\,l=f$ & 109.2812 & --4.7143 & 8.4 \\
   &                &            &          &       \\
SiO       &        $J=2-1$   &    86.8470   & --4.5335  &	6.3 \\
          &        $J=5-4$\tablefootmark{a}   &    217.1050   & --3.2843	&   31.3 \\
   &                &            &          &       \\
$^{29}$SiO &        $J=2-1$   &    85.7592   & --4.5500	&   6.2 \\
   &                &            &          &       \\
SiS        &        $J=5-4$   &    90.7716  & --4.9241	&  13.1  \\
           &        $J=6-5$   &    108.9243 & --4.6799	&  18.3  \\
   &                &            &          &       \\
SO\tablefootmark{b}         &        $N=2-1,J=2-1$  &    86.0940  & --5.2799	& 19.3  \\
       &        $N=3-2,J=2-1$  &    99.2999  & --4.9488	&  9.2  \\
       &        $N=4-4,J=5-4$  &    100.0296 & --5.9656	& 38.6  \\
       &        $N=2-1,J=3-2$  &    109.2522 & --4.9665	& 21.1  \\
   &                &            &          &       \\
$^{34}$SO\tablefootmark{b}   &        $N=2-1,J=2-1$  &  84.4107  &  --5.30558 &  19.2 \\
	   &        $N=4-4,J=5-4$  &  96.7818  &  --6.0061 &  38.1 \\
	   &        $N=3-2,J=2-1$  &  97.7153  &  --4.9695 &  9.1  \\
	   &        $N=2-1,J=3-2$  &  106.7432  &  --4.9970 &  20.9 \\
   &                &            &          &       \\
$^{33}$SO\tablefootmark{c}  &        $N=2-1,J=2-1$  &     85.2379 &  --5.2928  &	19.3 \\
   &        $N=3-2,J=2-1$  & 	98.4936  &  --4.9591 &  9.1  \\
  &                &            &          &       \\
SO$_2$\tablefootmark{d}  &  $J_{K_a,K_b}=3_{1,3}-2_{0,2}$  &  104.0294 &  --4.2264  &	7.7 \\
        & $J_{K_a,K_b}=16_{2,14}-15_{3,13}$   & 104.0336  & --5.4987  &  138  \\ 
        &  $J_{K_a,K_b}=10_{1,9}-10_{0,10}$  &  104.2393 &  --3.7708 &	55  \\
\hline
    \end{tabular}
    \tablefoot{All parameters are taken from the Cologne Database for Molecular Spectroscopy (CDMS\footnote{https://cdms.astro.uni-koeln.de/cdms/portal/}; \citealt{Endres16})
    \tablefoottext{a}{From observations in \citet{Beltran22};}
    \tablefoottext{b}{For simplicity, in the text and figures we will label the SO and $^{34}$SO transitions only with the $N$ quantum number. We add the $J$ of the upper level for the two transitions having the same $N$ quantum numbers;}
    \tablefoottext{c}{We list only the two (expected-to-be) strongest hyperfine transitions, $F=7/2-5/2$ and $F=9/2-7/2$.}
    \tablefoottext{d}{We list only the lines detected towards the PN emitting regions. }
    }
    \label{tab:1}
\end{table}

\section{Observations and data reduction}
\label{obs}

Observations  towards  the  HMC  G31  were  taken  with  ALMA during Cycle 5 (project 2017.1.00501.S, P.I.: M.T. Beltr\'an) obtaining an unbiased spectral survey in Band 3 from 84.05 GHz up to 115.91 GHz. The frequency resolution is 0.49 MHz,  corresponding  to  a  velocity  resolution  of $\sim$1.6 \kms\
at  90~GHz.  The  final  angular  resolution  is 1\pas2 ($\sim$4500 au). 
The primary beam is $\sim 68$\asec\ at 84~GHz and $\sim 50$\asec\ at 115~GHz.
The pointing centre of the observations is R.A.(J2000)=18$^{\rm h}$47$^{\rm m}$34\pas 312 and
Dec.(J2000)=$-$01$^{\circ}$12$^{\prime}$45\pas 9. The uncertainty on the flux calibration is $\sim 5\%$. 
For more details on the data reduction (calibration, baseline subtraction, cleaning and line identification) we refer to \citet{Mininni20}.
The spectral analysis has been performed with the MAdrid Data CUBe Analysis 
({\sc madcuba}\footnote{{\sc madcuba} is a software developed in the Madrid Center of Astro-biology (INTA-CSIC) which enables to visualise and analyse single spectra and data cubes: https://cab.inta-csic.es/madcuba/.}, Mart\'in et al.~\citeyear{Martin19}) software, and will be described in Sect.~\ref{column}.

\section{Results}
\label{res}

\subsection{Emission morphology}
\label{morphology}

\subsubsection{PN, PO and SiO}
\label{maps-PN}

   \begin{figure*}
   \centering
   \includegraphics[width=0.74\hsize]{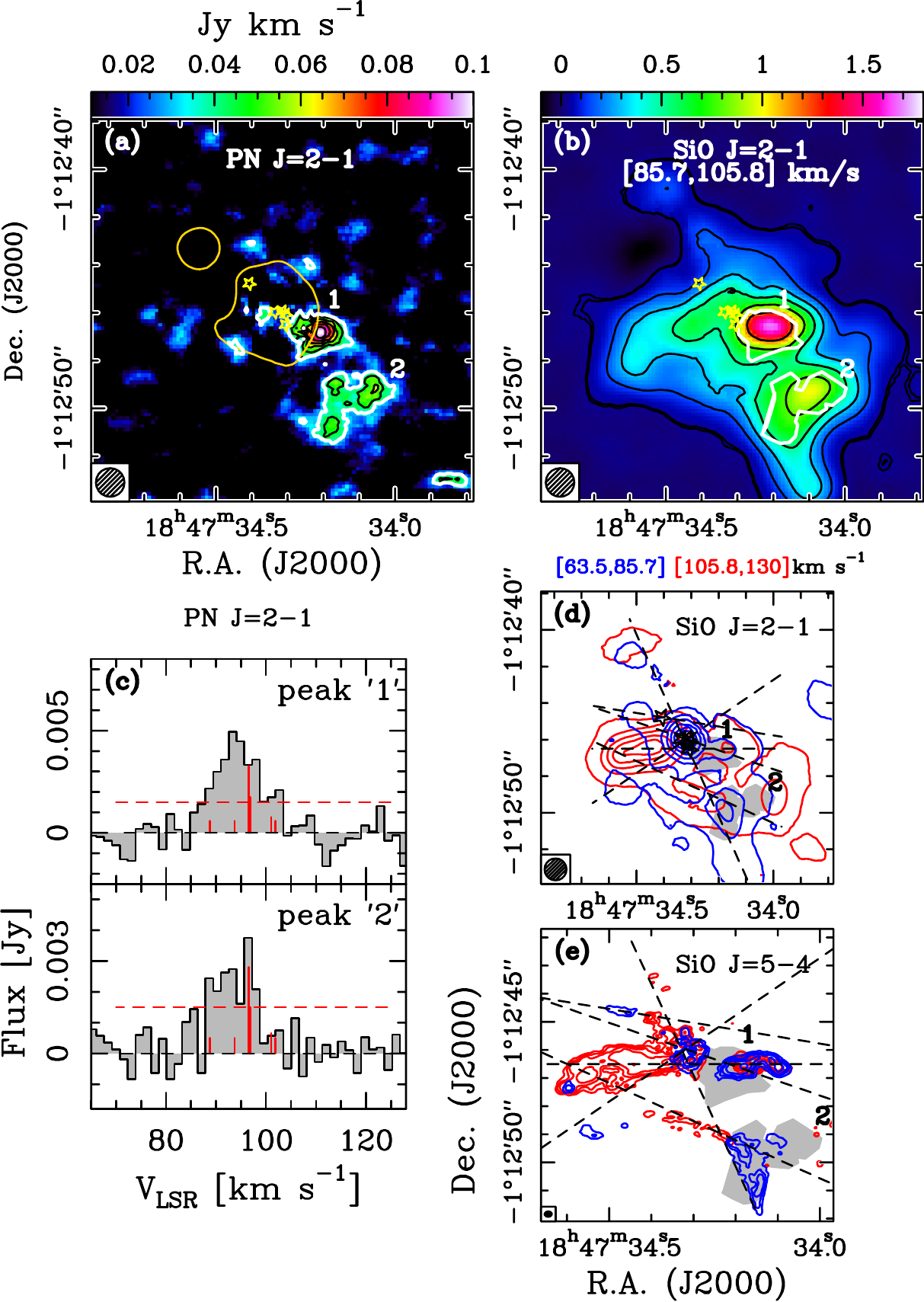}
      \caption{Intensity maps of PN and SiO integrated in velocity.
      \newline
      {\em (a):} PN $J=2-1$ integrated in the range 85.7--105.8~\kms. The white contour is the $3\sigma$ rms level of the integrated map ($\sigma = 1.14\times 10^{-2}$ Jy~\kms), while the black ones are in steps of 1$\sigma$ rms. The numbered regions 1 and 2 are those used to extract the spectra of all species analysed. The synthesised beam is in the bottom-left corner.
      The yellow contour is the 3~mm continuum emission at 16 mJy beam$^{-1}$, corresponding to 20$\sigma$ rms. 
      The yellow stars indicate the continuum sources identified by \citet{Beltran21}.
      \newline
           {\em (b):} map of the intensity of SiO $J=2-1$ integrated in the same velocity range as PN (colour scale). The PN emission regions identified in panel (a) are highlighted in white.
           Contours start at the 3$\sigma$ rms level of the integrated emission ($3\times 10^{-2}$~Jy \kms), and correspond to 3, 15, 30, 50, 80, and 120$\sigma$. 
           \newline
           {\em (c):} Spectra of PN $J=2-1$ extracted from the emission peak in regions "1" (upper panel) and "2" (lower panel). 
           The red dashed horizontal line indicates the 3$\sigma$ rms level, and the vertical lines show the position in velocity of the line hyperfine components.
           \newline
           {\em (d):} SiO $J=2-1$ emission integrated in the velocity ranges 63.5 -- 85.7~\kms\ (blue contours) and 105.8 -- 130.0~\kms\ (red contours). In both cases, the starting contour is the 3$\sigma$ rms level of the integrated map ($1.95\times 10^{-2}$ Jy \kms\ for the red contours, $1.62\times 10^{-2}$ Jy \kms\ for the blue contours), and the step is 25$\sigma$. The dashed lines correspond to the six outflows identified by \citet{Beltran22} from SiO $J=5-4$.
           The grey filled areas correspond to the PN emitting regions 1 and 2 identified in panel (a).
           \newline
           {\em (e):} same as panel (d) for SiO $J=5-4$, obtained at an angular resolution of $\sim 0.22$\asec\ with ALMA, and already published in \citet{Beltran18}. The integration velocity intervals are the same as in panel (d).}
         \label{fig:PN-integ}
   \end{figure*}

Figure~\ref{fig:PN-integ} shows the map of PN $J=2-1$ (panel (a)) integrated in the velocity range with emission above $3\sigma$~rms, namely 85.7--105.8~\kms. 
As reference, the systemic velocity of G31 is 96.5~\kms.
The emission morphology is extended and located mostly towards two regions, both south of the hot core peak. 
The emission peak of the most intense one is offset by --1\pas 6,--0\pas 9 from the phase centre, corresponding to $\sim 6900$ au. 
It is labelled as "1" in Fig.~\ref{fig:PN-integ}. 
The second one has a complex shape with multiple peaks. 
The main intensity peak is offset by --3\pas 9,--3\pas3 (corresponding to $\sim 19000$ au) from the phase centre, and is labelled region "2" in Fig.~\ref{fig:PN-integ}.
Towards both peaks, we extracted the PN $J=2-1$ spectra, which are shown in Fig.~\ref{fig:PN-integ}. The intensity of
both lines is clearly above the 3$\sigma$ rms level.
PO is only tentatively detected towards G31, so that we do not show the integrated map here, and we refer to Sect.~\ref{po} for the analysis of this tentative detection.

Figure~\ref{fig:PN-integ} also shows the emission of SiO $J=2-1$ integrated in the same velocity range as PN $J=2-1$ (85.7 -- 105.8~\kms, panel (b)). 
The emission morphology is in very good agreement with that of PN. 
In particular, SiO $J=2-1$ does not peak on the hot core but towards the two PN emission regions. 
Evidence that PN and SiO are spatially associated was already found from both single-dish studies \citep{Mininni18,Rivilla18,Fontani19,Lefloch16}, and interferometric studies \citep{Bergner22}. The maps in Fig.~\ref{fig:PN-integ} confirm very clearly this association.
The PN emission is less extended than that of SiO, perhaps due to insufficient sensitivity. 
To quantify this, first we have derived the contours at half maximum of the emission of PN and SiO in region "1".
The contours subtend solid angles with equivalent diameters of $\sim 3$\asec\ and $\sim 4$\asec, respectively. 
Hence, they are quite comparable.
Second, we can quantify what the expected emission of PN would be in the envelope detected in SiO scaling the SiO integrated intensity by the PN/SiO abundance ratio.
The average SiO integrated intensity in the envelope surrounding the emission peak "1" is $\sim 0.4$ Jy \kms. 
Multiplying this value by the lowest expected PN/SiO relative abundance, which is $\sim 0.1$ as we will show in Sect.~\ref{discu:abundances}, the integrated emission of PN would be $\sim 4\times 10^{-2}$ Jy \kms, marginally higher than the 3$\sigma$ rms level in the PN map that is $3.5\times 10^{-2}$ Jy \kms, and hence consistent with a non detection in PN.

Moreover, SiO is detected also at higher blue- and red-shifted velocities with respect to the velocities detected in PN.
We show the integrated emission in the blue and red wings (63.5--87~\kms\ and 102--130~\kms, respectively) in panel (c) of Fig.~\ref{fig:PN-integ}. 
PN is undetected at these high velocities. 
However, the remarkable agreement between PN and SiO in the velocity range in which PN emission is significant, and where SiO emission is most intense, suggests that the two molecules have very similar emission structure. 

In panel (d) of Fig.~\ref{fig:PN-integ} we show the maps of the high-velocity wings of the SiO $J=5-4$ line published in \citet{Beltran18}. 
These maps have been obtained at an angular resolution of $\sim 0.22$\asec, that is $\sim 5$ times better than that of the GUAPOS ones. 
The orientation of the six outflows in the region and the position of their driving sources has been recently improved with ALMA ($\sim 0.09$\asec, Beltr\'an et al.~\citeyear{Beltran22}). 
We can see that the PN emission is entirely to the SW of the sources driving the outflows, and it can be associated with four of the six flows identified in \citet{Beltran22} at higher angular resolution. 
Thus, the complex PN emission we see in both regions "1" and "2" is probably due to the superposition of several outflow lobes. 
The superposition also suggests that PN seems to be mainly associated with the blue-shifted SiO outflow lobes (panel (d) in Fig.~\ref{fig:PN-integ}). In fact, even though region "1" is associated with both red-shifted and blue-shifted emission, region "2" is only detected in the blue-shifted emission.
This could be due to the blue-shifted lobes being brighter compared to the red ones. 
Other reasons, such as inhomogeneity of the dense gas in the clump where the hot core (and outflows) are embedded, are also possible.

\begin{figure}
   \centering
   {\includegraphics[width=0.9\hsize]{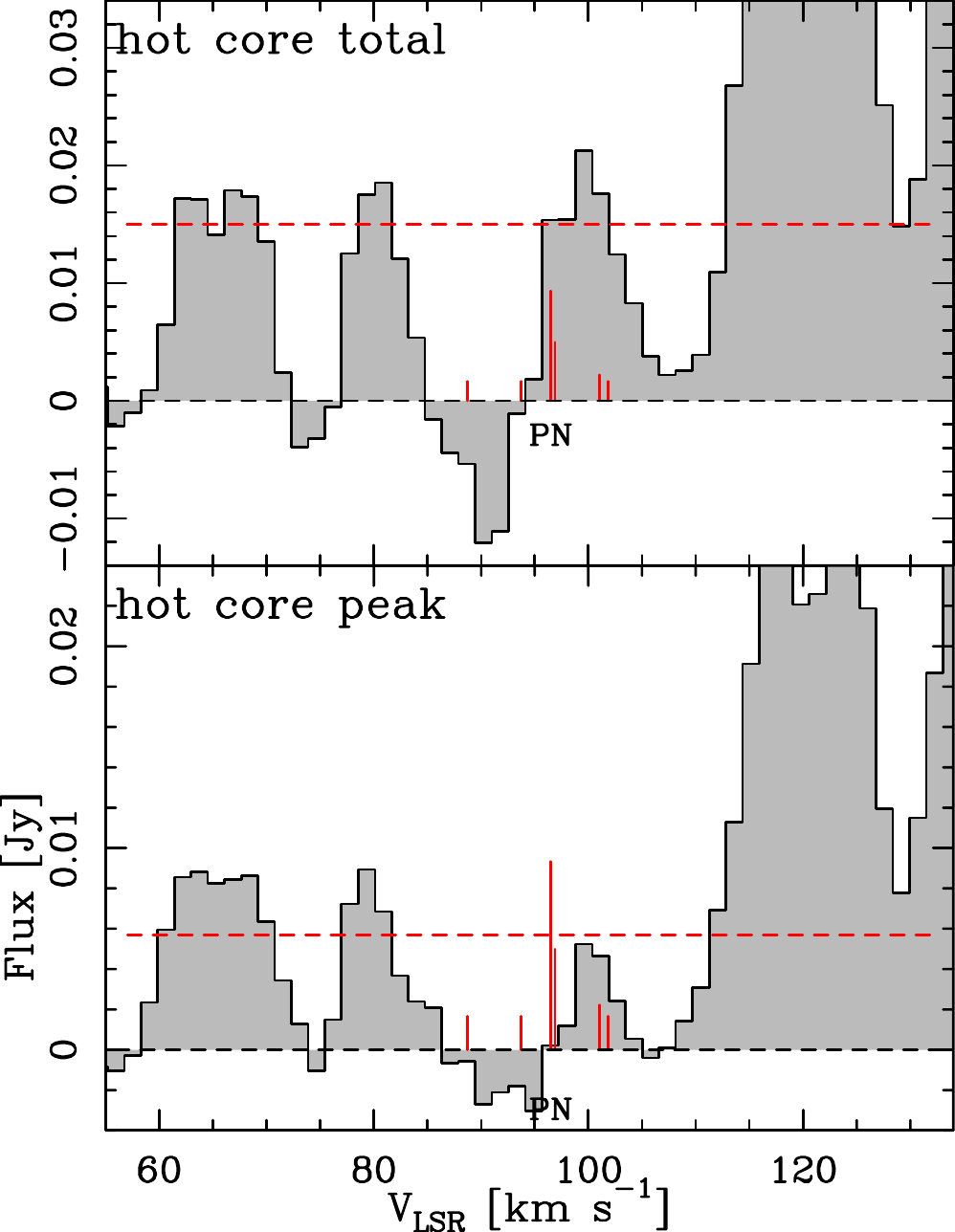}
   }
   \caption{Integrated PN $J=2-1$ spectrum extracted from the hot core.
   The upper panel shows the spectrum extracted from the 3~mm continuum contour shown in Fig.~\ref{fig:PN-integ}. 
   The lower panel shows the spectrum extracted from a beam around the intensity peak of the 3~mm continuum. 
   The red vertical lines indicate the expected velocities of the hyperfine components, and their length is proportional to the relative intensity of the components.
   The horizontal lines illustrate the 3$\sigma$ rms level.}
   \label{fig:PN-hotcore}
\end{figure}

The PN emission is also clearly off the hot core location, and this result is consistent with what has been observed with ALMA in the young stellar objects AFGL5142 and B1 \citep{Rivilla20,Bergner22}. 
We checked if some PN emission arises also from the hot core analysing the spectrum extracted towards the 3~mm continuum \citep{Mininni20}. 
The spectrum extracted from the 3~mm continuum contour surrounding the hot core (Fig.~\ref{fig:PN-integ}) is shown in Fig.~\ref{fig:PN-hotcore}. 
We chose the 3~mm continuum contour at 20$\sigma$ rms to disentangle the emission of the hot core from that of the ultracompact H{\sc ii} region placed NE of the hot core \citep{Mininni20}.
We also show the spectrum extracted from a beam centred on the continuum intensity peak of the hot core.
We detect a faint emission at $\sim 100$~\kms\ and $\sim 80$~\kms\, hence not centred on the systemic velocity of G31, which is $96.5$~\kms.
The feature at $\sim 100$~\kms\ is unlikely to be a nearby line because no other species have been found to emit lines at the frequency of this feature
in previous GUAPOS works \citep{Mininni20,Colzi21,LopezGallifa23}.
Moreover, it is narrower (full width at half maximum $\sim 6$~\kms) with respect to the lines detected towards the hot core, which have width at half maximum broader than 7~\kms\ \citep{Mininni20,Mininni23,Colzi21}.
The hint of an absorption feature nearby (at $\sim 90$\kms) suggests that this could be PN partially self-absorbed.
The fact that this absorption and emission features are both more prominent towards the total hot core than towards the peak could indicate that it is PN emission from the envelope, possibly associated with narrower features, and/or partially self-absorbed in the blue part of the line.
However, the candidate absorption feature is too close to the noise level in both spectra, and 
hence we conclude that the emission from the hot core is lacking or negligible, as also suggested in \citet{LopezGallifa23}. 
The feature at $\sim 80$~\kms\ could be due to n-C$_3$H$_7$CN $32_{5,28} - 32_{4,23}$ at 93.9839~GHz and/or gGg'-(CH$_2$OH)$_2$ $15_{1,14}-14_{2,12}$ at 93.9818~GHz.

We stress that towards region "2" we identified all lines in the full GUAPOS spectrum, and found no lines of other identified species that could contaminate the PN $J=2-1$ line. 
Moreover, all lines detected towards this region were identified. 
Towards region "1" the identification of all species is more difficult because the full spectrum is very rich of lines.
However, the clean detection towards region "2" and the lack
of candidate contaminating lines from other species towards the hot core (Fig. 2) makes a contamination from other lines in region "1" also unlikely.
The identification of all species in regions "1" and "2", and their comparison, goes beyond the scope of this paper and will be the subject of a forthcoming work.

\subsubsection{SO, SiS, and SO$_2$}
\label{maps-SO}

\begin{figure*}
   \centering
   \includegraphics[width=1.\hsize]{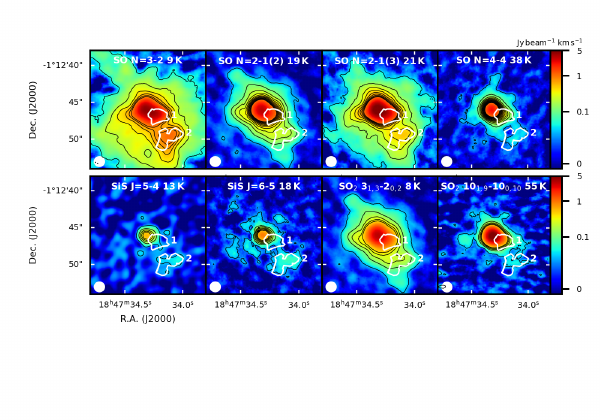}\vspace{-2.7cm}
      \caption{Emission of SO, SiS, and SO$_2$ integrated in velocity.
      {\em Upper panels:} velocity-integrated emission of, from left to right, SO $N=3-2$, $N=2-1(2)$, $N=2-1(3)$, and $N=4-4$ (see Table~\ref{tab:1} for the spectral parameters). The integration velocity range is 85.7--105.8~\kms\ in all images to match the velocity interval where the PN emission is detected. The black contours start from the $3\sigma$ rms level of the integrated maps, which is, from left to right: $4.8\times 10^{-2}$ Jy beam$^{-1}$ \kms; $2.5\times 10^{-2}$ Jy beam$^{-1}$ \kms; $3.6\times 10^{-2}$ Jy beam$^{-1}$ \kms;
      $2.7\times 10^{-2}$ Jy beam$^{-1}$ \kms, and the step is of $10\sigma$ rms. 
      The white contour corresponds to the PN integrated emission, and the numbered regions "1" and "2" are those defined in Fig.~\ref{fig:PN-integ}. The synthesised beam is illustrated in the lower-left corner.
      To the {\rm right} of the quantum numbers, we indicate the $E_{\rm up}$ of the transition.
      {\em Lower panels:} same as the upper panels for the SiS $J=5-4$ and $J=6-5$ lines, and the SO$_2$ $J_{K_a,K_b}=3_{1,3}-2_{0,2}$ and $J_{K_a,K_b}=10_{1,9}-10_{0,10}$ lines. 
      The 3$\sigma$ rms level is: $2.9\times 10^{-2}$ Jy beam$^{-1}$ \kms\ for SiS $J=5-4$; $2.3\times 10^{-2}$ Jy beam$^{-1}$ \kms\ for SiS $J=6-5$; $2.95\times 10^{-3}$ Jy beam$^{-1}$ \kms\ for SO$_2$ $J_{K_a,K_b}=3_{1,3}-3_{0,2}$;  $1\times 10^{-2}$ Jy beam$^{-1}$ \kms\ for SO$_2$ $J_{K_a,K_b}=10_{1,9}-10_{0,10}$. 
      }
         \label{fig:SOSiS-integ}
   \end{figure*}

Figure~\ref{fig:SOSiS-integ} shows the integrated intensity maps of the SO, SiS, and SO$_2$ transitions observed in GUAPOS and detected towards the PN emitting regions. 
The excitation analysis of all transitions of these species detected towards the hot core is performed in \citet{LopezGallifa23}.
The SO emission arises mostly from the hot core, and shows some extended emission to the south-west, with a secondary peak towards the PN emitting region "2" clearly visible in transitions $N=3-2$ and $N=2-1(3)$. 
Hence, overall the bulk of SO emission has a different morphology with respect to both PN and SiO.
The emission of both SiS and SO$_2$ is also dominated by the hot core, but there is a clear secondary peak coincident with region "2".
The increase of the SO, SiS, and SO$_2$ integrated intensity towards the hot core is naturally explained if warm/hot gas-phase chemistry is boosting the formation of sulphur-bearing species.
However, the SiS lines are also likely contaminated.
In fact, SiS $J=5-4$ is partly blended in the hot core with a line at 90.7697~GHz, unidentified so far, and the SiS $J=6-5$ is blended with CH$_3$COCH$_3$ at 108.9237 GHz and to an unidentified line detected in the hot core at 108.9216 GHz.
We will describe in Sect.~\ref{spectra} how such contaminations are clear in region "1", closer to the hot core, and disappear in region "2".
 
\begin{figure*}
   \centering
   {\includegraphics[width=0.9\hsize]{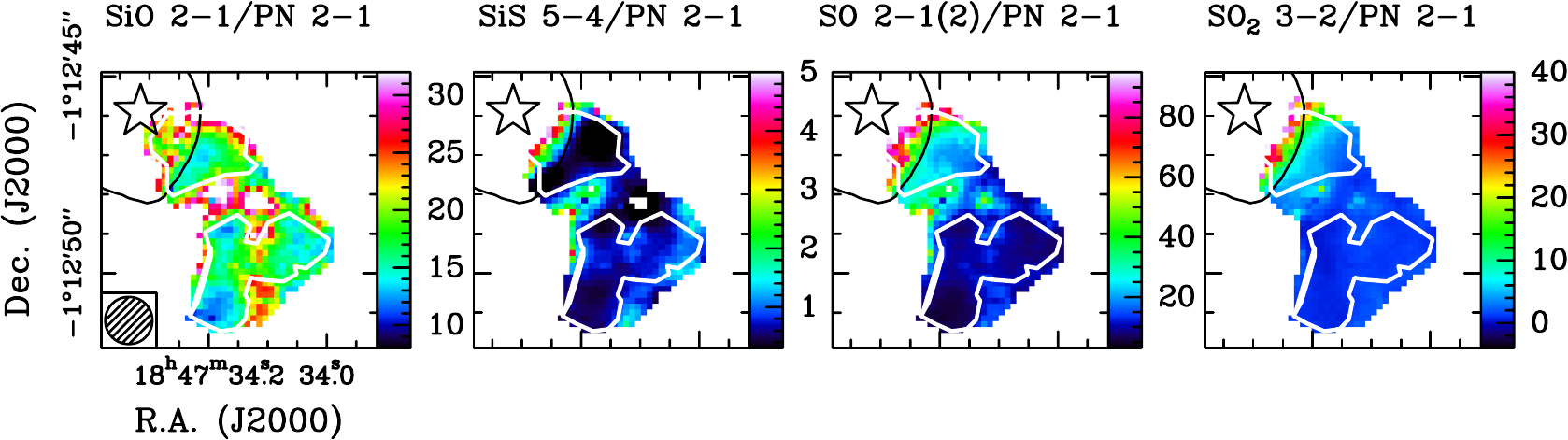}}
      \caption{Maps of integrated intensity ratios. The plots show the ratio between the PN integrated intensity and those of, from left to right: SiO $J=2-1$, SiS $J=5-4$, SO $J=2-1(2)$, and SO$_2$ $J_{K_a,K_b}=3_{1,3}-2_{0,2}$.
      We show only the emission included, or close to, the two PN emitting regions "1" and "2", as the PN integrated emission goes rapidly to zero elsewhere.
      The star in the top left corner indicates the phase center, and the ellipse in the bottom left corner in the first panel is the GUAPOS synthesised beam.
      The black and white contours correspond to the 3~mm continuum emission and the PN integrated emission, respectively, as shown in Fig.~\ref{fig:PN-integ}.}
         \label{fig:ratios-fig4}
   \end{figure*}

In summary, from the integrated intensity maps of all species the SiO emission is the most similar to the PN one. 
This is further illustrated in Fig.~\ref{fig:ratios-fig4}, where we plot the pixel-by-pixel ratio between the PN integrated intensity map and those of SiO $J=2-1$, SiS $J=5-4$, SO $N=2-1(2)$, and SO$_2$ $J_{K_a,K_b}=3_{1,3}-2_{0,2}$.
The ratio is nearly constant (within a factor two) for SiO/PN in both regions "1" and "2", while region "1" shows an increase up to a factor 10 of the ratios SiS/PN, SO/PN, and SO$_2$/PN towards the centre of the hot core. 
This shows further that both PN and SiO are formed predominantly in the outflows or in their cavities, while the Sulphur-bearing species are produced mostly in the hot core.

\subsubsection{$^{29}$SiO and $^{34}$SO}
\label{iso}

\begin{figure*}
   \centering
   {\includegraphics[width=1.\hsize]{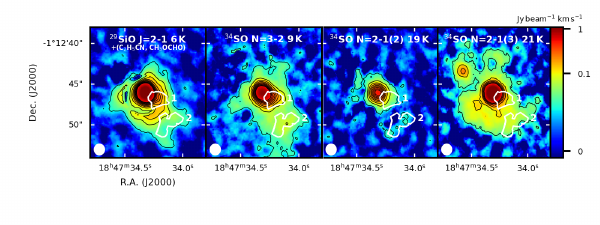}}\vspace{-1.2cm}
    \caption{Velocity-integrated emission of less abundant isotopologues.
    We show, from left to right, $^{29}$SiO $J=2-1$, $^{34}$SO $N=3-2$, $N=2-1(2)$, and $N=2-1(3)$. 
    The integration velocity range is 85--101~\kms\ for $^{29}$SiO, which is the range in which PN and $^{29}$SiO are both detected. 
    For $^{34}$SO, we have used: 81--105 for the $N=3-2$ line, 88--104 \kms\ for the $N=2-1(2)$ line, and 85.7--105.8~\kms\ for the $N=2-1(3)$ line. 
    The black contours start at the $3\sigma$ rms level of the integrated intensity maps, which is, from left to right: $1.6\times 10^{-2}$ Jy beam$^{-1}$ \kms; $1.8\times 10^{-2}$ Jy beam$^{-1}$ \kms;
    $1.7\times 10^{-2}$ Jy beam$^{-1}$ \kms; $8.7\times 10^{-3}$ Jy beam$^{-1}$ \kms, and the step is $10\sigma$ rms. 
   The white contour corresponds to the $3\sigma$ rms level of the integrated map of PN, and the numbered regions "1" and "2" are those defined in Fig.~\ref{fig:PN-integ}. The synthesised beam is illustrated in the bottom-left corner.
      }
         \label{fig:34SO-integ}
   \end{figure*}

Figure~\ref{fig:34SO-integ} shows the integrated intensity maps of the lines of the less abundant isotopologues clearly detected.
Because these are expected to be optically thin, they should illustrate the actual morphology of the molecular emission better than their main isotopologues.
The $^{29}$SiO emission peaks on the hot core, contrary to the main isotopologue. 
However, this line is likely contaminated by two transitions: CH$_3$OCHO $21_{5,16}-21_{4,17}$ at 85.761876~GHz, and C$_2$H$_5$CN $J_{(K_1,K_2)}=11_{2,10}-11_{1,11}$ at 85.760502~GHz. The centroid velocity of this line is expected to be displaced by about $-4.5$~\kms\ from $^{29}$SiO $J=2-1$.
Because both CH$_3$OCHO and C$_2$H$_5$CN peak towards the hot core as the other COMs detected in the region, the fact that the emission peak in the map is on the hot core is likely due to this contamination.

The emission morphologies of the $^{34}$SO lines are very similar to those of their main isotopologues: the $N=3-2$ and $N=2-1(3)$ transitions both peak towards the hot core but have a clear secondary peak towards the PN region "2". 
The $N=2-1(2)$ line is mostly concentrated on the hot core and the secondary peak is barely visible, perhaps due to insufficient sensitivity. 
Therefore, the maps of the less abundant isotopologues of SO confirm the morphology of the main one, while that of SiO is different due to blending with a line arising from the hot core.
We stress, however, that a possible partial contamination in region "1" is possible also in the $^{34}$SO lines. In particular, the $N=2-1(2)$ line could be contaminated by nearby lines of CH$_3$COOH and CH$_3$COCH$_3$ detected in the hot core.
This would explain why the map of $^{34}$SO in this line is more concentrated towards the hot core than the others.
The $N=3-2$ and $N=2-1(3)$ could also be contaminated by nearby lines detected but unidentified in the hot core.
Hence, in summary the analysis of all these lines towards region "1" should be regarded with caution due to strong contaminations.

\subsection{Spectra}
\label{spectra}

\subsubsection{Main isotopologues}
\label{main-iso}

\begin{figure}
   \centering
   \includegraphics[width=1.\hsize]{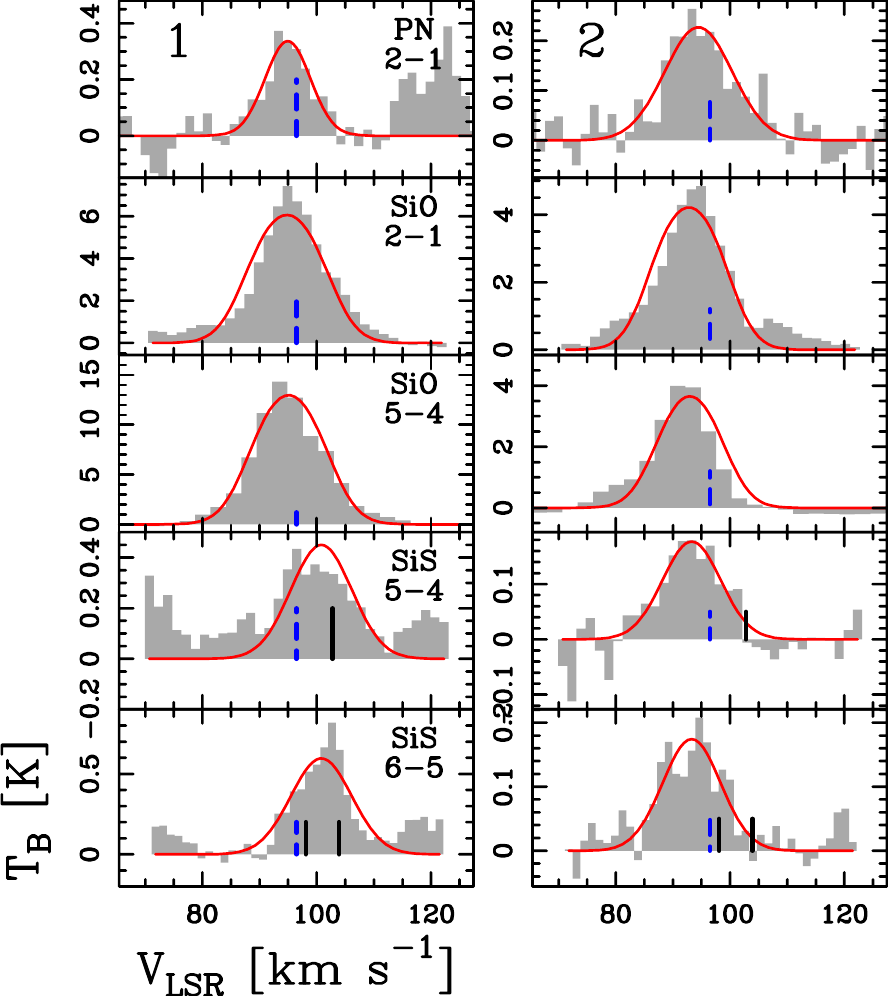}
      \caption{Spectra extracted from the regions detected in PN $J=2-1$. 
      We show the transitions (from top to bottom) of PN ($J=2-1$), SiO ($J=2-1$ and $J=5-4$), and SiS ($J=5-4$ and $J=6-5$).
      On top of each column we give the number of the extraction region according to the labels in Fig.~\ref{fig:PN-integ}. 
      The blue dashed vertical lines correspond to the reference LSR velocity ($96.5$~\kms, Mininni et al.~\citeyear{Mininni20}). 
      The black vertical lines in the panels of the SiS $J=5-4$ line correspond to an unidentified line detected in the hot core at 90.7697~GHz, and in those of the $J=6-5$ transition correspond to CH$_3$COCH$_3$ at 108.9237~GHz and to an unidentified line at 108.9216~GHz, both detected in the hot core \citep{Mininni23}.
      The red curves superimposed on the lines represent the best LTE fit of the analysed molecules obtained with {\sc madcuba}.
      For PN, we show the fit obtained fixing \Tex\ to that of SiO, that is 26~K (see Sect.~\ref{coldens-pn}).
              }
         \label{fig:spectra-1}
   \end{figure}

\begin{figure}
   \centering
   \includegraphics[width=1.\hsize]{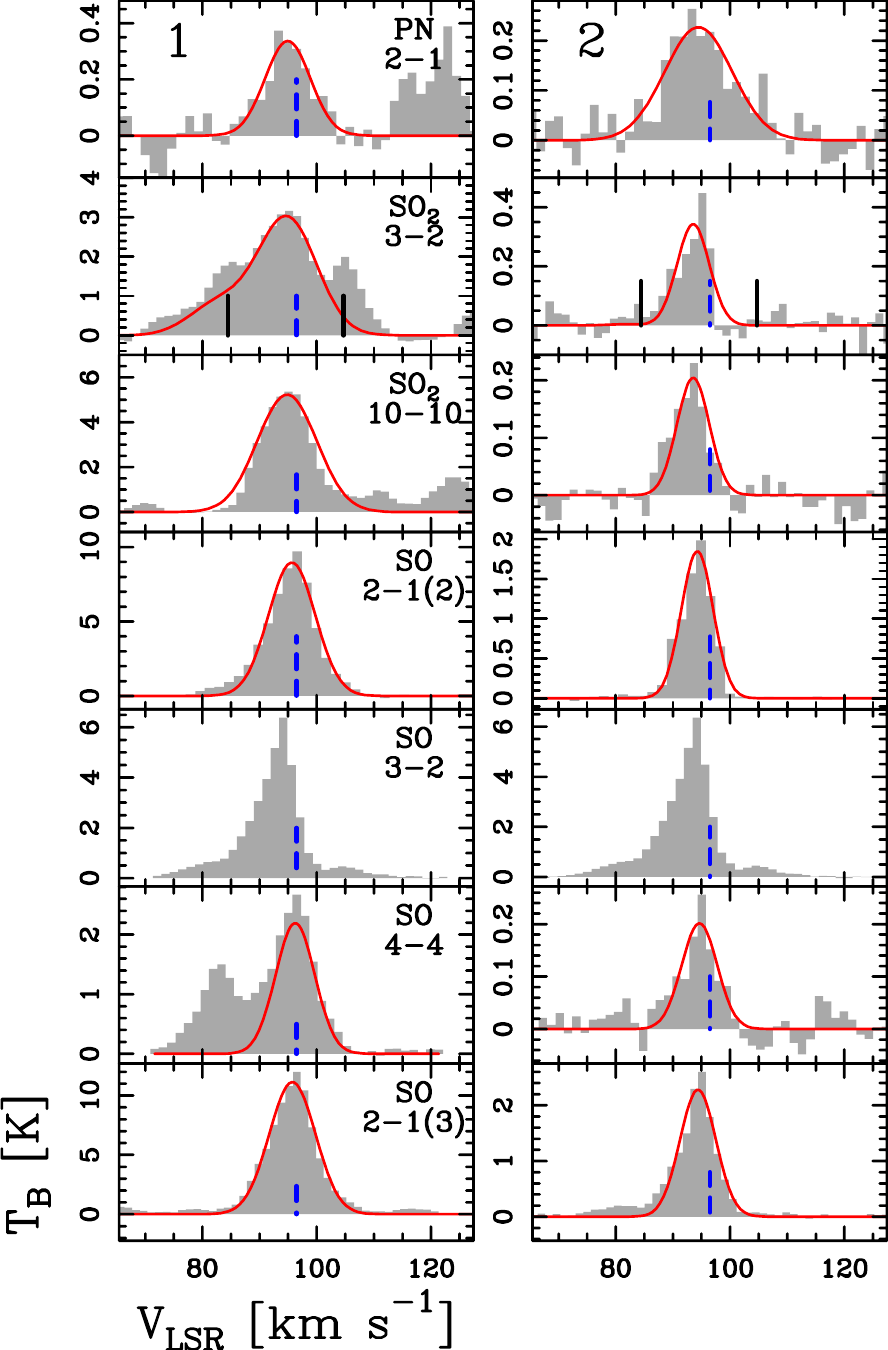}
      \caption{Same as Fig.~\ref{fig:spectra-1}, showing the spectra of SO$_2$ ($J=3-2$ and $J=10-10$), and SO ($J=2-1$, $J=3-2$, and $J=4-4$). 
      We keep as reference the spectra of PN in the top panels as in Fig.~\ref{fig:spectra-1}.
      The black vertical lines in the spectrum of SO$_2$ $J=3-2$ are SO$_2$ $J=16-15$ (Table~\ref{tab:1}) and $^{13}$CH$_3$OH $J=13-14$ at 104.0266~GHz.}
         \label{fig:spectra-so-so2}
   \end{figure}

In Figures~\ref{fig:spectra-1} and \ref{fig:spectra-so-so2} we show the spectra of the transitions of the main isotopologues reported in Table~\ref{tab:1}, extracted at the two emitting regions of PN $J=2-1$ illustrated in Fig.~\ref{fig:PN-integ}.
The spectra are in brightness temperature units.
The conversion from flux density units to temperature units has been performed using Eq.(1) in \citet{Fontani21},
which provides an average brightness temperature over the angular surface of each region. 

\begin{figure*}
   \centering
   \includegraphics[width=0.8\hsize]{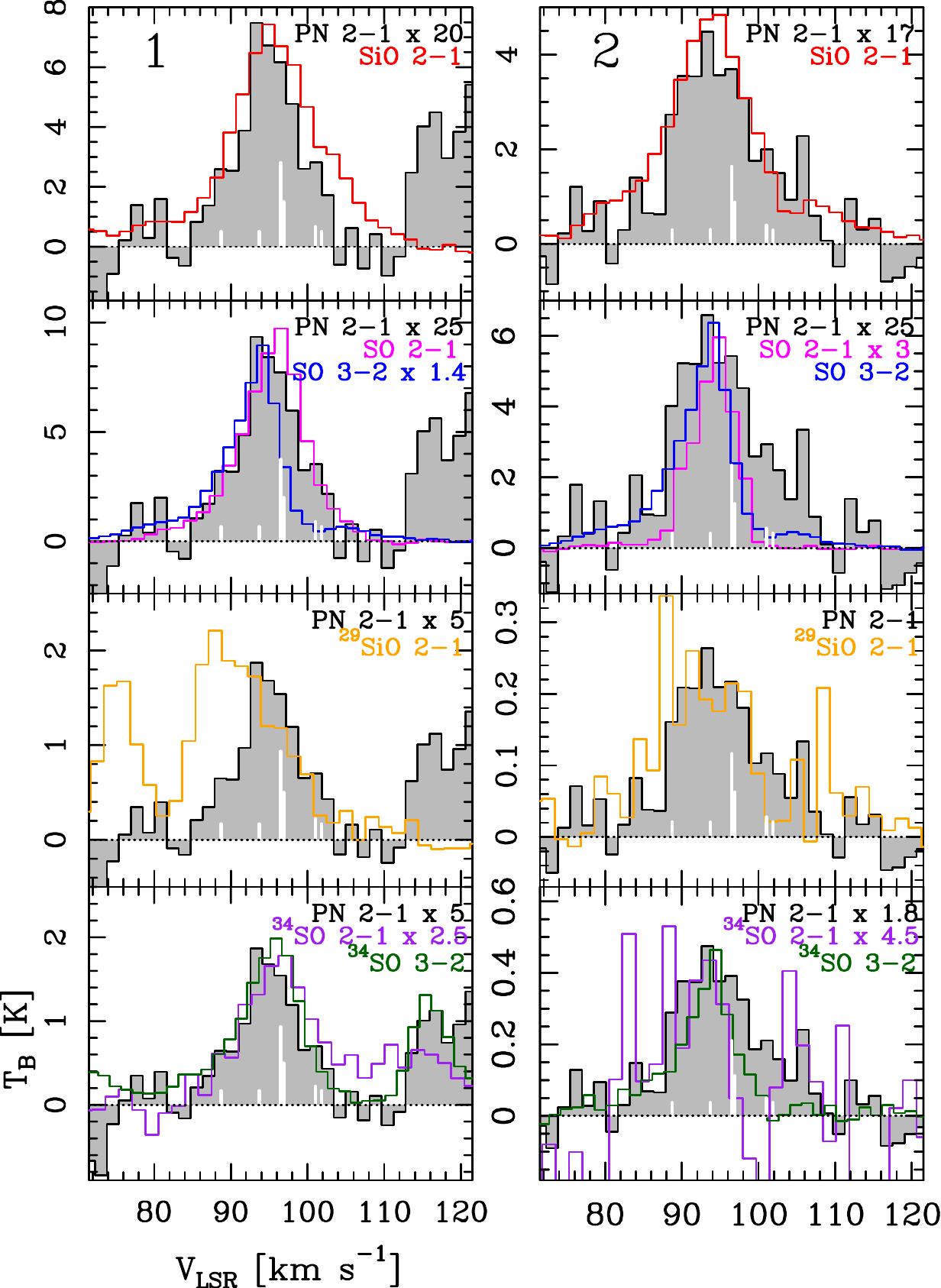}
   \caption{Spectral comparison between PN, SiO and SO.
   The PN $J=2-1$ spectra (grey histogram) extracted from regions "1" and "2" in Fig.~\ref{fig:PN-integ}, are superimposed on SiO $J=2-1$ (red histograms), SO $N=2-1(2)$ (magenta histograms), SO $N=3-2$ (blue histograms), $^{29}$SiO $J=2-1$ (orange histograms), $^{34}$SO $N=2-1(2)$ (purple histograms), and $^{34}$SO $N=3-2$ (green histograms) extracted from the same regions. 
   The PN line intensity scale has been appropriately multiplied to perform a consistent spectral comparison with SiO and SO. 
   In each plot, the white vertical bars are the theoretical positions in velocity of the PN $J=2-1$ hyperfine components (Table~\ref{tab:1}), with the strongest one being at the systemic velocity of $96.5$~\kms.}
        \label{fig:spectra-overlaps}
   \end{figure*}

The spectral profiles of SiO and PN are compared in Fig.~\ref{fig:spectra-overlaps}, and appear similar in both regions "1" and "2". This similarity is especially apparent in the central velocity channels, while it is less apparent at high velocities ($\geq 102$ \kms, and $\leq 87$ \kms), where the wings detected in SiO disappear in PN. 
This is likely due to insufficient sensitivity in this velocity range for PN. The detection of PN in a narrower velocity range with respect to SiO has also been found through ALMA observations in the high-mass protostar AFGL5142 \citep{Rivilla20} and the low-mass protostar B1-a \citep{Bergner22}, as well as in single-dish studies \citep{Lefloch16,Mininni18,Rivilla18,Fontani19}.

Both SiS lines appear displaced (red-shifted) in velocity towards region "1" with respect to PN, SiO, and SO (Fig.~\ref{fig:spectra-1}). 
Such velocity shift is not seen in region "2", where the SiS velocity peak in both lines is placed very close to that of PN, SiO, and SO.
We think that, in region "1", the SiS $J=5-4$ line is contaminated by an unidentified line clearly detected towards the hot core at 90.7697~GHz, and the SiS $J=6-5$ line is contaminated by emission of CH$_3$COCH$_3$ at 108.9237 GHz arising from the hot core, and by
an unidentified line at 108.9216~GHz, both clearly detected towards the hot core \citep{Mininni23,Colzi21}.

The profiles of the SO transitions $N=2-1(2)$ and $N=3-2$ are, overall, similar to PN in region "1" (Fig.~\ref{fig:spectra-so-so2}), even though the intensity peaks are slightly different: PN $J=2-1$ and SO $N=3-2$ peak at $\sim 94$ \kms, while SO $N=2-1(2)$ peaks at the G31 systemic velocity, that is $96.5$~\kms. 
In region "2", the SO lines are narrower than those of PN and SiO, suggesting that SO can be dominated by more quiescent material, maybe surrounding the outflow lobes, towards this position rather than by shocked material. 

Finally, the SO$_2$ lines have a profile similar to that of SO in region "2", while in region "1" there are some differences: the $J_{K_a,K_b}=10_{1,9}-10_{0,10}$ transition has a profile similar to SO, while the $3_{1,3}-2_{0,2}$ transition shows a blended profile with the SO$_2$ $16_{2,14}-15_{3,13}$ line (undetected in region "2") and
with $^{13}$CH$_3$OH $13_{-3,11}-14_{-2,13}$ at 104.02655~GHz, detected towards the hot core \citep{Mininni23} and not towards region "2".

\subsubsection{Tentative detection of PO}
\label{po}

\begin{figure*}
   \centering
   \includegraphics[width=0.9\hsize]{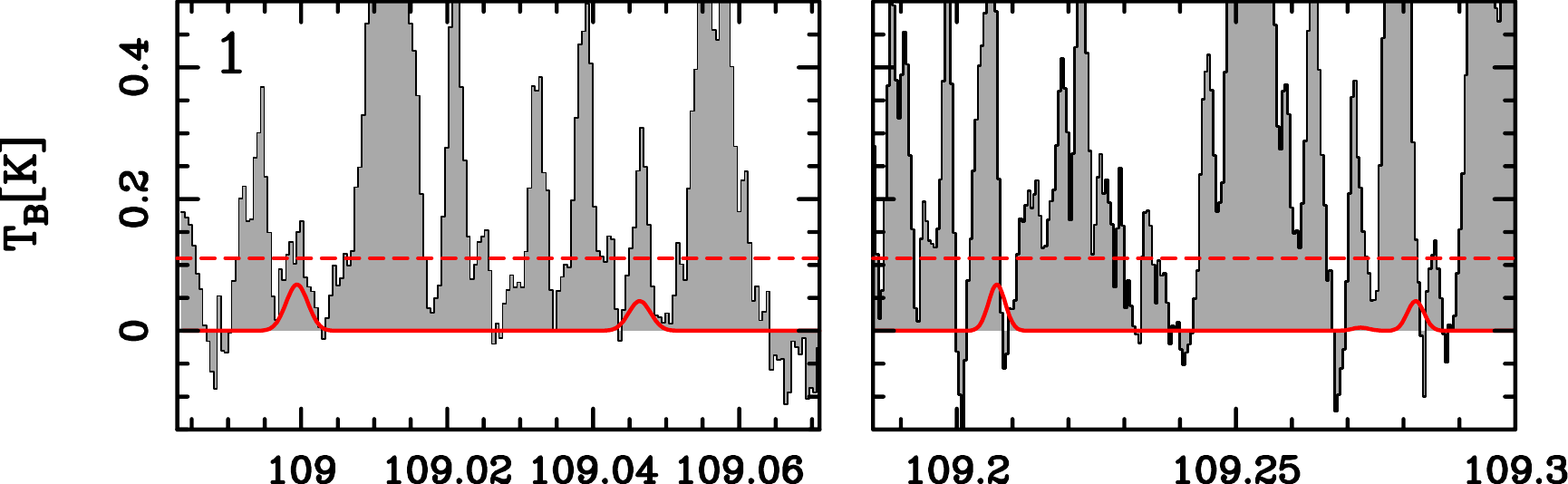}
   \vskip0.3cm
   \includegraphics[width=0.9\hsize]{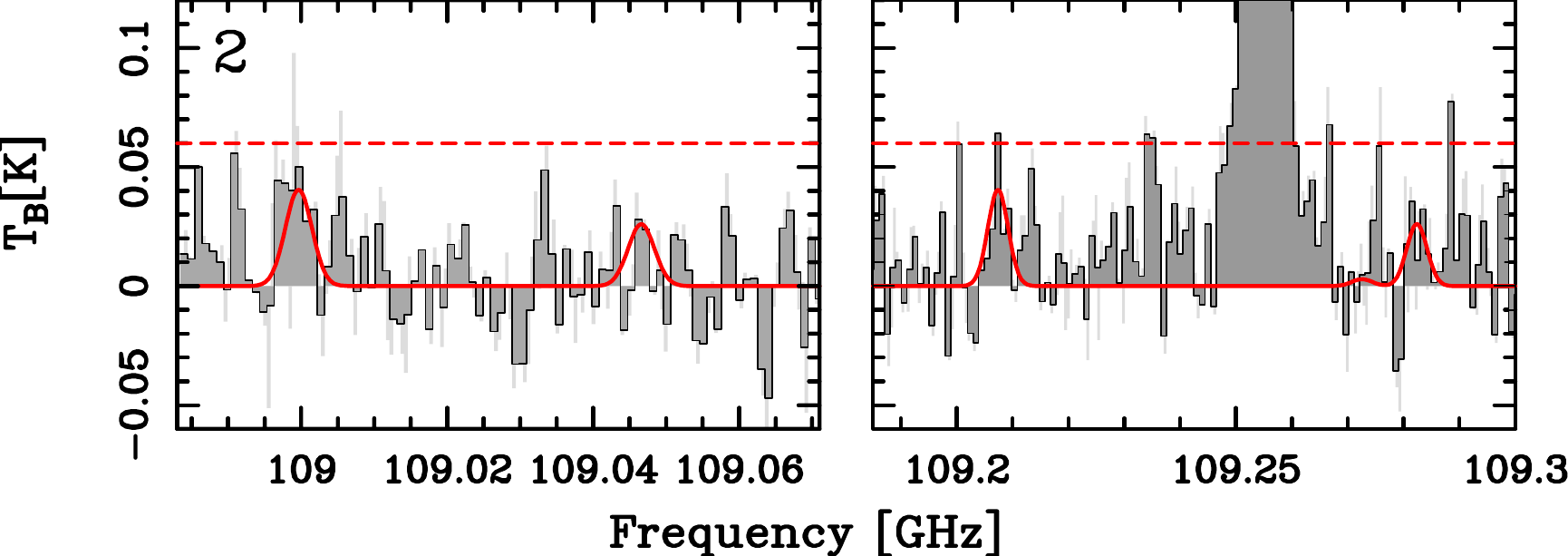}
   \caption{Spectra containing the PO lines in the GUAPOS observations. 
   {\em Upper panel:} spectrum extracted towards region "1". 
   The red curve indicates the PO lines fitted according to the best-fit parameters in Table~\ref{tab:fit-total} (assuming $T_{\rm ex}$=26 K).
   The horizontal red dashed line corresponds to the $3\sigma$~rms level.
    {\em Lower panel:} same as top panel for region "2", using the fit shown in Table~\ref{tab:fit-total} assuming  $T_{\rm ex}$=12 K. The observed spectra at original frequency resolution are shown in light gray, while the spectra smoothed by a factor of 2 are shown in dark gray.}
        \label{fig:spectra-po}
   \end{figure*}

The four lines of PO listed in Table~\ref{tab:1} extracted from regions "1" and "2" are shown in the spectra of Fig.~\ref{fig:spectra-po}.
We used {\sc madcuba} to fit simultaneously the four transitions.
The fitting method, which assumes local thermodynamic equilibrium (LTE) conditions, is fully described in Sect.~\ref{coldens-so}.
We have used the spectroscopic entry 047507 of the Cologne Database for Molecular Spectroscopy (CDMS), based on the laboratory measurements of
\citet{Bailleux02} and \citet{Kawaguchi83}, and the dipole moment determined by \citet{Kanata88}.

In region "1" (upper panel in Fig.~\ref{fig:spectra-po}), a detection of PO cannot be ruled out but it is difficult to claim due to the spectral complexity and the contamination by emission lines from other molecules (Fig.~\ref{fig:spectra-po}).
In particular, the two PO transitions in the right panel of Fig.~\ref{fig:spectra-po}, namely the transitions at 109.2062 GHz and 109.2812 GHz, are blended with transitions of ethyl formate (C$_2$H$_5$OCHO) and ethylene glycol (aGg-CH$_2$OH)$_2$, respectively, as already found in source W51  \citep{Rivilla16}.
The lines contaminating the PO transitions in the left panel are not obvious to identify, as a proper determination of all contaminating species would require the study of the full GUAPOS spectrum towards this position. 
This goes beyond the scope of this paper and will be the subject of a forthcoming work.

In region "2" (lower panel in Fig.~\ref{fig:spectra-po}), less affected by nearby lines, the synthetic PO spectrum is consistent with a tentative detection. 
If we consider the peak intensity, the four transitions are below the $3\sigma$~rms noise, but
if we consider the signal-to-noise ratio on the line integrated intensity, the two strongest transitions, that are at 108.9984~GHz and 109.2062~GHz, are consistent with a detection. 
The signal to noise ratio on the integrated intensity can be computed from the expression $\int T_{\rm B} {\rm d}V  / [\sigma \times \sqrt{\Delta V/FWHM} \times FWHM]$, where $\Delta V$ is the spectral resolution in velocity.
Plugging in this expression the line integrated intensity (0.51 K \kms), the $1\sigma$~rms (0.02~K), and the FWHM obtained with {\sc madcuba} (12~\kms, see Table~\ref{tab:fit-total}), we obtain a signal-to-noise ratio of $\sim 6$.
The spectrum towards this position was smoothed in velocity by a factor 2 to increase the sensitivity. 
The tentative detection is supported by the fact that the emission is most likely detected towards the two lines expected to be the strongest in the quadruplet (i.e. at 108.9984~GHz and 109.2062~GHz).

\subsubsection{Less abundant isotopologues}
\label{less-iso}

\begin{figure}
   \centering
   \includegraphics[width=1.\hsize]{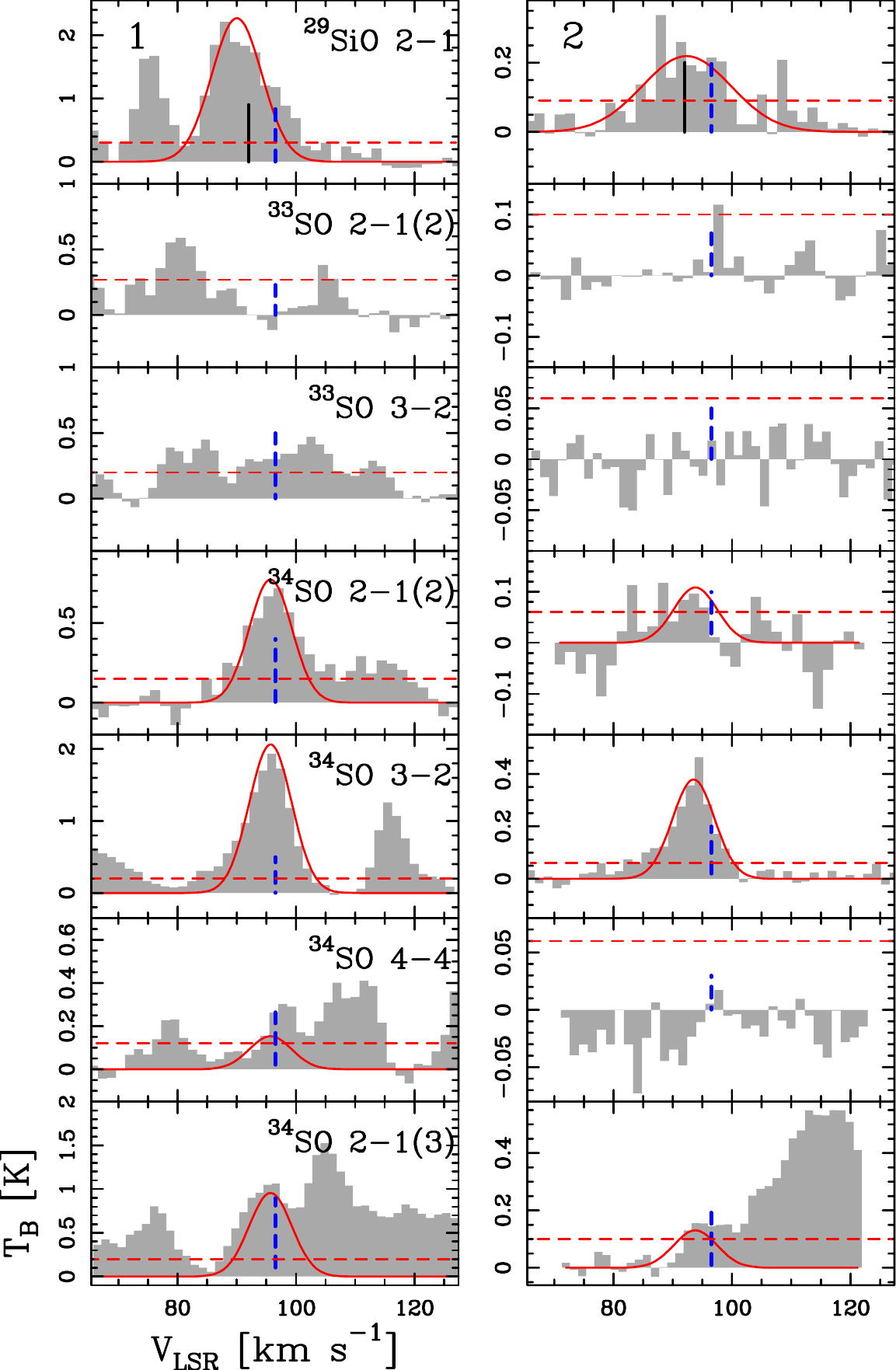}
      \caption{Same as Fig.~\ref{fig:spectra-1} for the lines of $^{29}$SiO $J=2-1$, $^{33}$SO $N=2-1(2)$ and $N=3-2$, and $^{34}$SO $N=2-1(2)$, $N=3-2$, $N=4-4$, and $N=2-1(3)$.
      The black solid vertical line in the spectrum of $^{29}$SiO $J=2-1$ indicates the expected peak velocity of C$_2$H$_5$CN $J_{K_a,K_b}=11_{2,10}-11_{1,11}$.
      In each panel, the horizontal red dashed line shows the 3$\sigma$ rms level.}    
         \label{fig:spectra-2}
   \end{figure}

The spectra of the less abundant isotopologues, namely $^{29}$SiO, $^{34}$SO, and $^{33}$SO 
extracted from regions "1" and "2" are shown in Fig.~\ref{fig:spectra-2}.
We detected clearly $^{29}$SiO $J=2-1$, $^{34}$SO $N=2-1(1), 3-2$ and $4-4$, in both regions. 
The detection of $^{34}$SO $N=2-1(2)$ and $4-4$ in region "2" is likely but difficult to firmly confirm because it is very close to the noise level.
The line profiles of the $^{34}$SO lines are overall similar in both line width at half maximum and centroid velocity to the corresponding $^{32}$SO ones (compare Figs.~\ref{fig:spectra-1} and \ref{fig:spectra-2}). 

The line profile of $^{29}$SiO $J=2-1$ is similar to the $^{28}$SiO one in the red wing, but relatively much brighter than the $^{28}$SiO one in the blue wing. As already noticed in Sect.~\ref{maps-PN}, this excess emission is likely due to contamination from C$_2$H$_5$CN emission, the expected centroid velocity of which is indicated in Fig.~\ref{fig:spectra-2}.

Figure~\ref{fig:spectra-overlaps} shows the PN $J=2-1$ line profile superimposed on the detected lines of $^{29}$SiO and $^{34}$SO.
For $^{34}$SO, we use the two unblended lines $N=2-1(2)$ and $N=3-2$.
In region "2", the PN and $^{29}$SiO line profiles are very similar since $^{29}$SiO is likely less blended.
The $^{34}$SO lines are very similar to PN in region "1", and narrower in region "2". As already noticed for the main isotopologue SO lines, this could be due to the fact that the SO emission is associated with a more quiescent material towards this position.

We did not clearly detect any $^{33}$SO line, even though towards region "1" a tentative detection of the $J=3-2$ transition is possible, as shown in Fig.~\ref{fig:spectra-2}. 
Assuming that the line is optically thin in both less abundant isotopologues, we can check if the tentative detection in $^{33}$SO $N=3-2$ is consistent with the expected line intensity: towards region "1" the $^{34}$S/$^{33}$S ratio is $\sim 5$. Because the relative isotopic ratio $^{34}$S/$^{33}$S is $\sim 5.6$ \citep{Lodders03}, the measured ratio is
consistent with a tentative detection, but the $^{33}$SO line also suffers from severe blending with nearby transitions, and hence we will not analyse it further.

\subsection{Molecular column densities and excitation temperatures}
\label{column}

In this section we derive the molecular column densities, \Ntot, of the analysed species. We derive first \Ntot\ and excitation temperature, \Tex, of the molecules detected in multiple lines, namely SO, $^{34}$SO, SO$_2$, SiS, and SiO
(Sect~\ref{coldens-so}).
Then, using the excitation temperature derived from these species, we compute \Ntot\ of PN and PO (Sect~\ref{coldens-pn}).

\subsubsection{SO, $^{34}$SO, SO$_2$, SiS, and SiO}
\label{coldens-so}

The fit to the spectra of SO, $^{34}$SO, SO$_2$, SiS, and SiO shown in Figs.~\ref{fig:spectra-1}, \ref{fig:spectra-so-so2}, and \ref{fig:spectra-2}, and the derivation of the physical and spectral parameters has been performed with the Spectral Line Identification and LTE Modelling (SLIM) tool of {\sc madcuba}. Through its AUTOFIT function, SLIM produces the synthetic spectrum that best matches the data assuming LTE. The input parameters are: \Tex, \Ntot, radial velocity of the source ($V$), line full-width at half-maximum (FWHM), 
and angular size of the emission ($\theta_{\rm S}$). AUTOFIT assumes that $V$, FWHM, \Tex, and $\theta_{\rm S}$
are the same for all transitions fitted simultaneously. These input parameters have all been left free except $\theta_{\rm S}$, for which we can safely assume that the emission fills the telescope beam, as can be seen in Figs.~\ref{fig:SOSiS-integ} and \ref{fig:34SO-integ}.
The results are shown in Table~\ref{tab:fit-total}. 

Let us first present the best-fit parameters of the S-bearing species.
For $^{34}$SO, in region "1" the fit converged leaving $V$, $FWHM$, and \Ntot\ free, but we had to fix \Tex. 
The best-fit \Tex\ obtained by visual inspection is 58~K. 
The resulting column density in region "1" is $\sim 1.1\times 10^{15}$\cmq.
In region "2", the fit converged leaving all the parameters free, except the filling factor as explained above.
The resulting \Ntot\ is $\sim 8\times 10^{13}$\cmq, and \Tex\ is 25~K.
For SO, in region "1" the fit could not converge leaving all parameters free and using all lines. Hence, first we did not use the $N=3-2$ transition which has a spectral shape suggesting a (too) high optical depth. Then, we fitted the remaining three transitions fixing FWHM to 9~\kms, which corresponds to the best-fit value derived from $^{34}$SO. 
The resulting best-fit column density and \Tex\ are $1\times 10^{16}$~\cmq\ and 60~K, respectively.
In region "2" the fit converged leaving all the parameters free, and the resulting column density and \Tex\ are
$\sim 1.2\times 10^{15}$\cmq\ and 31~K, respectively.
The SO$_2$ excitation temperature in both regions is consistent within the uncertainties with the values measured from SO ($\sim 75$~K in region "1" and $\sim 27$~K in region "2"), and the column densities are $\sim 9.8\times 10^{15}$\cmq\ in region "1" and $\sim 1.6\times 10^{14}$\cmq\ in region "2".

The fit of the lines of the Si-bearing species SiS and SiO (Fig.\ref{fig:spectra-1}) converged leaving all parameters free except the filling factor in both regions. However, as discussed in Sect.~\ref{spectra}, the SiS lines in region "1" are both strongly blended with nearby lines, some of which are from unidentified species and hence could not be fitted simultaneously with SiS. Therefore, we derived \Ntot\ and \Tex\ only for region "2", where we obtain \Tex $\sim 13$~K and \Ntot $\sim 2.6\times 10^{13}$\cmq. 
For SiO, in region "1" we derive \Tex $\sim 26$~K and \Ntot $\sim 3.6\times 10^{14}$\cmq, and in region "2" \Tex $\sim 12$~K and \Ntot $\sim 1.8\times 10^{14}$\cmq.
Finally, we did not estimate the parameters for $^{29}$SiO because the line is too blended with a nearby line of C$_2$H$_5$CN, as discussed in Sect.~\ref{spectra}.

\begin{table*}
    \centering
    \caption{Best-fit spectral and physical parameters obtained with {\sc madcuba}.}
    \begin{tabular}{l l ccccc}
\hline
\hline
  region  &  molecule & FWHM  &   $V$  &   $N_{\rm tot}$ &  \Tex\  & $\tau_{\rm max}$\\
          &  &  (\kms ) &   (\kms )    &    (\cmq )        &  (K)   &  \\
          \hline
    "1"   & $^{34}$SO & 9.0(0.9)  & 95.7(0.4) & $1.1(0.1)\times 10^{15}$ &  58 & 0.04 \\
    &  SO & 9.0  &  96.0(0.5)   &    $1.0(0.2)\times 10^{16}$ &  60(10) & 0.34 \\
                &  SO$_2$ &  12.0(0.6) &  94.8(0.3)  & $9.8(1.5)\times 10^{15}$  &  75(10) & 0.08 \\
         &  SiS\tablefootmark{(a)}  & -- & -- & -- & -- \\
          &  SiO    &   12.6(0.4)  & 95.1(0.2)  &  $3.6(0.1)\times 10^{14}$  &  26.4(0.2) & 0.9 \\
          &  PN & 9.7(0.9)  &  94.8(0.4)   &    $1.3(0.1)\times 10^{13}$ &  26   &  0.01 \\
          &    &  9.7(0.9)  & 94.8(0.4)   & $3.1(0.3)\times 10^{13}$ & 75 & 0.01 \\
          &  PO & 9.7  & 94.8  & < 1.7 $\times$ 10$^{13}$ & 26  & -- \\
          &   &  9.7 & 94.8 & < 3.8 $\times$ 10$^{13}$ & 75 & -- \\          
          \hline
   "2"     & $^{34}$SO & 8.6(0.9) &  93.8(0.4) &    $8(2)\times 10^{13}$ & 25(10) & 0.02 \\
        & SO &  7.9(0.5) &  94.8(0.4)   &    $1.2(0.3)\times 10^{15}$        & 31(12) & 0.3 \\
        &  SO$_2$ & 6.2(0.6)  &  93.8(0.3) & $1.6(0.2)\times 10^{14}$  & 27(3) & 0.02 \\
        &  SiS  & 12.3(0.9)  & 93.3(0.4) & $2.6(0.3)\times 10^{13}$  &  13(1) & 0.02 \\
        &  SiO   &  11.5(0.6) &  93.0(0.3) & $1.8(0.2)\times 10^{14}$  &  11.6(0.6) & 0.7 \\
        &  PN &  11.7(1.2) &  94.2(0.5)   &    $1.5(0.1)\times 10^{13}$     &  12 & 0.01\\
        &     &   12.2(1.2)  &  94.3(0.5)  & $1.6(0.1)\times 10^{13}$  &   30 & 0.01 \\
        &  PO & 11.7 & 94.2 & $\sim$ 8.5(1.4) $\times$ 10$^{12}$ & 12 & 0.005 \\
             &  & 12.2 & 94.3  & $\sim$ 1.3(0.2) $\times$ 10$^{13}$ & 30 & < 0.001 \\
    \hline
    \end{tabular}
    \tablefoot{All parameters have been obtained with {\sc madcuba} as described in Sect.~\ref{coldens-so}. The numbers in brackets are the uncertainties. The quantities without uncertainties (in brackets) are fixed in the fit.
    \tablefoottext{a}{Both lines are strongly blended with nearby lines, some of which are from unidentified lines.}}
    \label{tab:fit-total}
\end{table*}

\subsubsection{PN and PO}
\label{coldens-pn}

For PN, we cannot derive \Tex\ from the data since we have only one transition.
Hence, we derive the best-fit in a range of \Tex\ based on the minimum and maximum values obtained from the other tracers. 
These are: 26--75~K in region "1", and 12--30~K in region "2". 
The results are given in Table~\ref{tab:fit-total}.
The PN total column densities are of the order of $10^{13}$~\cmq, regardless of the region and temperature assumed.
The largest PN column density is measured towards region "1" assuming \Tex=75~K.
We have fit the PN lines also taking the hyperfine structure into account, and derived in all cases FWHMs slightly smaller but consistent within the errors with the values obtained neglecting it.

For PO, as mentioned in Section \ref{po}, the PO transitions towards region "1" are significantly contaminated by brighter emission from other species. It was already noted by \citet{Rivilla16} that the transitions at 109.2062 GHz and 109.2812 GHz are blended with transitions of ethyl formate (C$_2$H$_5$OCHO) and ethylene glycol (aGg-CH$_2$OH)$_2$, respectively. 
The contamination of the other two lines is less obvious and would require the analysis of the full GUAPOS spectrum towards this position, which is beyond the scope of this paper. 
Since the presence of PO in this position cannot be confirmed, we have thus derived an upper limit for its column density. 
We have used the values of \Tex, FWHM and \VLSR  derived for PN, and using {\sc MADCUBA} we have increased the value of \Ntot to the maximum value that is consistent with the observing spectrum. The resulting LTE model is shown in Fig.  \ref{fig:spectra-po}, and the derived upper limit is $\sim 1.2-1.3$ (Table \ref{tab:fit-total}).

Regarding position "2", which is much less line-rich, the PO transitions are not contaminated, and they seem to be detected, especially the two brightest transitions. 
We have fitted them with {\sc MADCUBA}, using again the same \Tex, FWHM and \VLSR derived for PN. 
The resulting LTE fit and \Ntot\ are shown in Fig. \ref{fig:spectra-po}, and Table \ref{tab:fit-total}, respectively. 
The PO/PN ratio turns out to be $\leq 1$,
which is consistent with the results found in some low-mass star-forming regions \citep{WeB22}, but smaller than the typical PO/PN ratios measured in high-mass star-forming regions \citep{Rivilla16,Rivilla18} and in the comet 67P/Churyumov-Gerasimenko, where the PO/PN ratio is at least 10 \citep{Rivilla20}.

\subsubsection{Opacity of SiO and SO lines}
\label{opacities}

{\sc madcuba} estimates the line opacities as explained in \citet{Martin19}, and provides column densities already corrected for optical depth effects.
The optical depth at the line centroid velocity, $\tau_{\rm max}$, computed with {\sc madcuba} for each molecule is listed in Table~\ref{tab:fit-total}.
In case of multiple transitions, we list $\tau_{\rm max}$ of the line having the highest opacity.
The values shown in Table~\ref{tab:fit-total} indicate optically thin lines in most cases except SiO, for which $\tau_{\rm max}$ is around 1, and SO, for which $\tau_{\rm max}$ is $\sim 0.3$.

For both SiO and SO lines, the opacity at line peak can also be computed from the less abundant isotopologues. 
Some lines of these isotopologues have the same quantum numbers as those of the main one (see Table~\ref{tab:1}).
Hence, their excitation temperatures should also be comparable and the line opacity can be derived in an alternative way. If the emitting region is also the same for the two isotopologues, from the equation of radiative transfer one can demonstrate that the ratio between the line brightness temperature of two isotopologues depends only on their relative abundance, and on the optical depth of the main isotopologue. 
Let us consider the case of SiO: the line intensity ratio is given by
\begin{equation}
\frac{T_{\rm p}^{28}}{T_{\rm p}^{29}}\sim \frac{1-\exp^{-\tau_{28}}}{1-\exp^{-\tau_{28}/ X[28/29]}}\;,  
\label{eq:opacity}
\end{equation}
where $T_{\rm p}^{28}$ and $T_{\rm p}^{29}$ are the brightness temperatures of the two lines, $\tau_{28}$ is the optical depth of the main isotopologue, and $X[28/29]$ is the $^{28}$Si/$^{29}$Si relative abundance ratio.
If $\tau_{28}\ll 1$, Eq.(\ref{eq:opacity}) states that 
$\frac{T_{\rm p}^{28}}{T_{\rm p}^{29}}\sim  X[28/29]$, namely the temperature ratios should be equal to the expected isotopic ratio.
For SO, an equation similar to Eq.(\ref{eq:opacity}) is valid for the $^{32}$S/$^{34}$S and $^{33}$S/$^{34}$S ratios.

The reference Solar values for $^{28}$Si/$^{29}$Si and $^{32}$SO/$^{34}$SO are 19.7 \citep{AeG89} and 22.5 \citep{Lodders03}, respectively.
Inspection of Fig.~\ref{fig:spectra-overlaps} shows that, at 
the velocity where the $^{28}$SiO line peaks (i.e. $\sim 94.8$~\kms), the intensity ratio between the $^{28}$SiO and the $^{29}$SiO lines is $^{28}$SiO/$^{29}$SiO$\sim 8$ in region "1" and $^{28}$SiO/$^{29}$SiO$\sim 20$ in region "2". 
This indicates that in region "2", where the isotopic ratio is very close to the Solar one, both lines are likely optically thin.
In region "1" the computed $^{28}$SiO/$^{29}$SiO ratio gives $\tau_{28}\sim 2.35$, which is, however, an upper limit due to blending of $^{29}$SiO with C$_2$H$_5$CN. In fact, the value computed by {\sc madcuba} is $\tau_{28}\sim 1$.
In region "2" {\sc madcuba} provides $\tau_{28} \sim 0.7$, which is not consistent with optically thin lines, but at least with line opacity smaller than that in region "1".

The $^{32}$SO/$^{34}$SO ratio derived from the $N=2-1(1)$ line is $\sim 22$ in region "2", consistent with optically thin emission in both lines, and $^{32}$SO/$^{34}$SO$\sim 13$ in region "1", smaller than the reference value by a factor 1.7. 
This ratio provides an optical depth $\tau_{32}\sim 1.2$ larger than the 0.34 value provided by {\sc madcuba}, but still consistent with moderately optically thick lines.
As stated in Sect.~\ref{column}, the $N=3-2$ line is more optically thick in region "1", but it has been excluded from the fit due to its complex spectral shape that did not allow the fit to converge.

\section{Discussion}
\label{discu}

\subsection{Comparison of spatial emission between PN and shock tracers}
\label{discu:shock}

The most direct and apparent result of this study is the similar spatial distribution of PN and the SiO $J=2-1$ bulk-velocity emission.
This was already found in the intermediate-/high-mass protostellar object AFGL5142 \citep{Rivilla20}, as well as in the low-mass protostar B1-a \citep{Bergner22} with ALMA observations.
Our work agrees with the two mentioned works also in the fact that PN does not arise from the position of the protostar(s) embedded in the hot core.
The non detection of PN towards the hot core could be explained either by insufficient atomic P to form PN, which in turn is abundant along the outflow cavities owing to grain sputtering, or to disruption of the PN molecule in the high temperature and high irradiation environment of the hot core. 
However, other species sensitive to UV photodissociation, such as methanol, are abundant towards the hot core \citep{Mininni23}. 
\citet{Jimenez18} proposed that PH$_3$ is abundantly produced on grain surfaces via hydrogenation of P during the collapse phase. 
Then, upon evaporation, it is rapidly (in timescales of $10^4$~yrs) converted in PN and PO.
However, both molecules are not detected clearly in the hot core, and hence PH$_3$ does not seem to be the main P-carrier here. 
The fact that PN is detected only in the shocked regions indicates that the main carrier of P is in the dust cores, where sputtering is needed to destroy (partially or totally) the grains. 

The comparison between PN and SiS is less clear because the SiS emission is heavily contaminated by nearby lines close to the hot core.
However, the PN line studied here has an energy of the upper level lower than those of the SiS lines ($E_{\rm up}\sim 7$~K against $E_{\rm up}\sim 13$ and $\sim 18$~K, respectively), and hence could be associated with (slightly) different material.
If, and eventually how, the emission changes going to higher excitation lines needs to be investigated through maps of higher-$J$ lines of PN.
At present, the map of PN at the highest $J$ at high-angular resolution is the $J=3-2$ one in \citet{Bergner22}, performed with ALMA towards B1-a. 
This map looks very similar in morphology to the PN $J=2-1$ one obtained in the same work.
However, the upper energy level of PN $J=3-2$ is $\sim 13.5$~K, that is just a factor $\leq 2$ higher than the $J=2-1$ one. 
It would be hence interesting to map higher excitation PN lines to check if the intensity peak of the emission changes.

We also propose the PN emitting region "2" as a new "hot spot" for shocked material. Such region, well separated from the hot core, will allow us to study the chemistry of shocked gas without the influence of the hot core, similarly to other known chemically rich shocked protostellar spots like L1157--B1 \citep{Gueth98}.

\subsection{Comparison in velocity between PN and shock tracers}
\label{discu:shock}

Another aspect emerging from our observations is the presence of PN only at relatively low velocities if compared with the velocities attained in the wings of the SiO and SO emission (see e.g.~Fig.~\ref{fig:spectra-1}). 
This finding is, again, in agreement both with the high-angular resolution studies performed towards AFGL5241 and B1-a and with the single dish studies in \citet{Mininni18} and \citet{Fontani19}.
This could be due either to a lack of sensitivity in the high-velocity regime, or to the fact that PN is produced only in not-so-strong shocks. 
A similar behaviour is seen in young outflows (e.g.~L1448-mm) where molecules such as H$_2$S fade away rapidly for high velocities while SiO and SO remain very bright at all velocities. 
One possibility could be the destruction of the onion-shell structure of dust grains where volatile species are detected at lower velocities due to the erosion of the ices and SiO and SO are also seen at high velocities because they are released from the grain cores \citep{Jimenez05}. 
However, as discussed in Sect.~\ref{main-iso}, and also in \citet{Mininni18} and \citet{Fontani19}, the fact that the line profile of PN and SiO is similar in the low(er) velocity channels suggests that the non-detection of PN at high velocities is more likely due to the lack of sensitivity.

To quantitatively establish this, we computed what the intensity in the wings of the SiO lines would be if the SiO abundance would be that of PN.
The column density ratio PN/SiO is $\leq 1/10$ (Table~\ref{tab:fit-total}).
Fig.~\ref{fig:spectra-1} indicates that the maximum intensity in the high-velocity wings of SiO $J=2-1$ is $\sim 1.5$~K in region "1" and $\sim 1$~K in region "2".
Scaling down these values by the abundance ratio PN/SiO$\sim 1/10$, which is reasonable because the emission in the wings is optically thin, one would obtain maximum intensities of $0.15$~K and $0.1$~K, respectively. 
The $1\sigma$~rms noise in the SiO $J=2-1$ spectrum in region "1" is 0.1~K, and that in region "2" is 0.033~K.
Thus, the intensities in the SiO wings scaled down by the PN/SiO factor would be at most 1.5 and 3 times the rms.
Considering that these are upper limits, these results are consistent with a non-detection of the high-velocity wings, as we see in PN.

Concerning the PN peak velocity, in region "1" it is identical to the SiO one, and consistent within the uncertainties with those of SO, $^{34}$SO, and SO$_2$. In region "2" the PN peak velocity is consistent within the uncertainties with that of all the other tracers except SiO, but the difference between the two values is just $\sim 1.2-1.3$~\kms, comparable to the velocity resolution of the observations.

\subsection{Column density comparisons}
\label{discu:abundances}

\begin{figure}
    \centering
    \includegraphics[width=1.0\hsize]{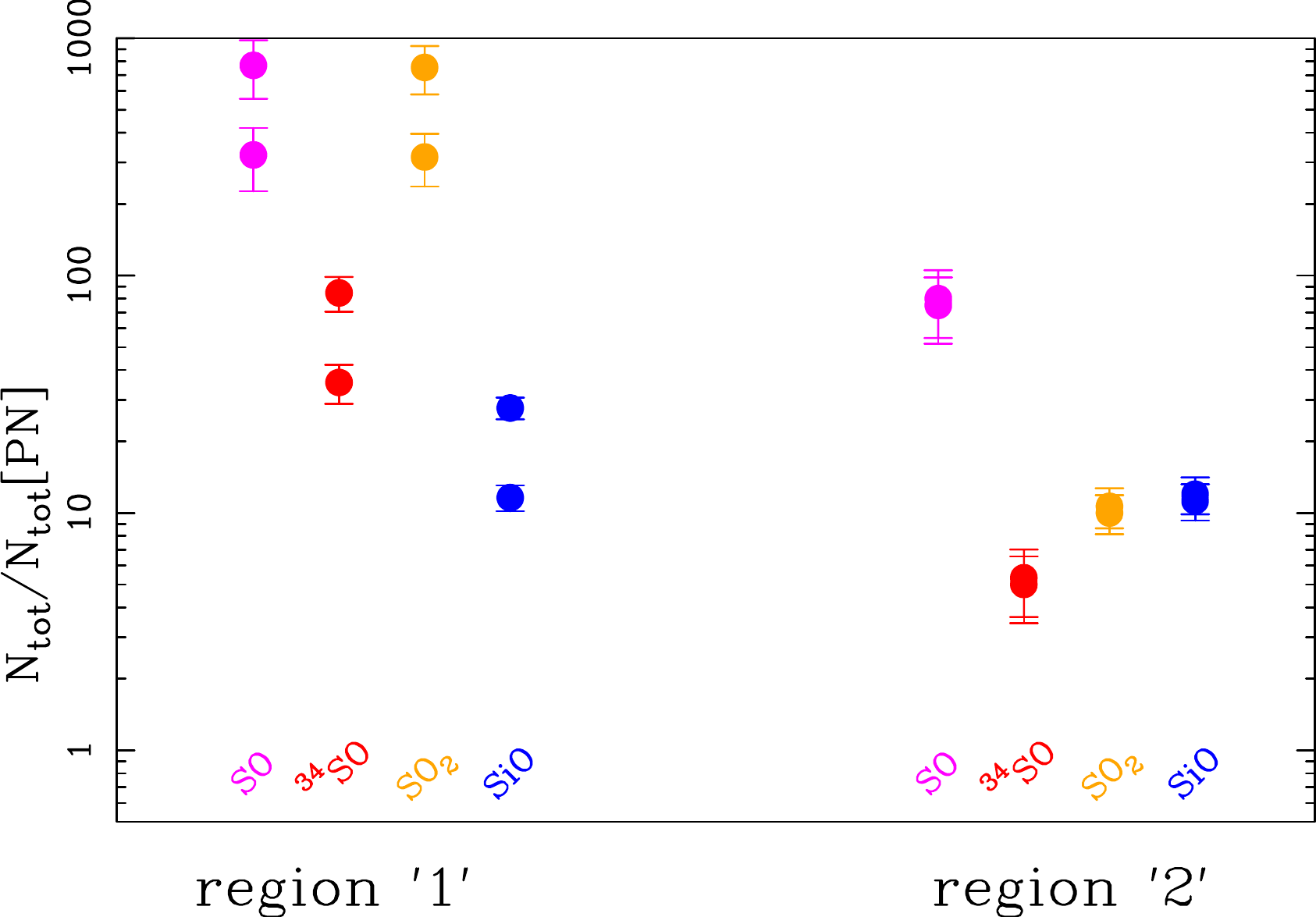}
    \caption{Column density ratios between molecular species and PN. 
    The different molecules are identified by different colours, and the measurements obtained in region "1" and "2" are separated on the x-axis as indicated. For each molecule, we plot two values of $N_{\rm tot}/N_{\rm tot}$[PN], which correspond to the two $N_{\rm tot}$[PN] obtained in the \Tex\ velocity interval given in Sect.~\ref{coldens-pn} (Table~\ref{tab:fit-total}).}
    \label{fig:ratios}
\end{figure}

Figure~\ref{fig:ratios} shows the molecular total column density ratios between PN and the other species studied in this work.
We plot the ratios for region "1" and "2" calculated assuming two \Tex\ as described in Sect.~\ref{coldens-pn}.
The SO/PN, $^{34}$SO/PN, and SO$_2$/PN ratios are higher in region "1" by about one order of magnitude than in region "2", while the SiO/PN ratios are consistent in the two regions within the uncertainties.
Because \Ntot\ of PN remains almost constant (around $10^{13}$~\cmq) in both "1" and "2", the different ratios between the two regions arise mostly from the decrease by an order of magnitude of \Ntot\ of SO, $^{34}$SO, and SiS. 
In Sect.~\ref{opacities}, we evaluated that the SiO $J=2-1$ line could be affected by optical depth towards region "1" possibly higher than that provided by {\sc madcuba}. The upper limit on $\tau$ estimated this way is $\sim 2.35$, instead of $\sim 0.9$ obtained with {\sc madcuba}.
Correcting \Ntot\ in Table~\ref{tab:fit-total} for such different $\tau$, one would obtain towards region "1" \Ntot $\sim 6 \times 10^{14}$~\cmq, and the SiO/PN ratios would become 66 and 24 for \Tex[PN] = 12 and 60~K, respectively.
Even in this case the SiO/PN ratio for \Tex[PN] = 12 would be marginally consistent with the value in region "2".

The SO/PN ratio was found to vary by orders of magnitude in the protostar AFGL5142 \citep{Rivilla20}. In the latter source, a high-mass protostar is driving a bipolar high-velocity jet surrounded by a cavity, both clearly detected in SO.
Several emission spots of PN and PO were detected along the cavity walls, but they were both undetected towards the protostar and the jet. 
Towards the protostar and the jet, the SO/PN ratio was of the order of 1000 or more, similar to the value obtained in G31 towards region "1".
Along the outflow {\it cavities}, instead, the SO/PN ratio drops down to $\sim 70-200$, consistent with the values we measure in region "2".
As discussed in Sect.~\ref{opacities}, the SO lines are affected by non-negligible opacities. In such case, the derived total column densities of SO could be lower limits, and so the SO/PN column density ratios should be even higher.
The $^{34}$SO/PN should not be affected by high optical depth effects, and even in this case the column density ratio drops by about an order of magnitude from region "1" to region "2".
In summary, our study confirms that PN ans SiO are very selective tracers of outflow cavities, unlike SO and SO$_2$.

Finally, the PO/PN ratio in region "2", where PO is tentatively detected, is $\sim 0.6-0.9$, and 
in region "1", the PO/PN upper limit is $\sim 1.2-1.3$, depending on the assumed temperature.
These ratios, even though in line with previous measurements on low- and high-mass protostars \citep{Rivilla16,Lefloch16,WeB22}, are lower than those obtained in the outflow spots of AFGL5142 \citep{Rivilla20}, where PO/PN$\geq 1$ and increasing with the distance from the protostar.
Ratios PO/PN$\leq 1$ are found in the theoretical models of \citet{Jimenez18} in case of pure shock models, without the need of a high cosmic ray ionisation rate, which in turn would be needed to explain PO/PN $> 1$.


\section{Conclusions}
\label{conc}

In the context of the GUAPOS project, we have studied P-bearing molecules towards the HMC G31 at high-angular resolution to investigate their connection with shock chemistry.
\begin{itemize}
\item We have clearly detected the PN $J=2-1$ transition and several SO, SO$_2$, SiO, and SiS rotational lines. PO lines are tentatively detected.
\item
The integrated intensity maps indicate that the emission of PN arises from two regions, "1" and "2", both southwest of the hot core peak, where four of the six outflows detected in \citet{Beltran22} are placed. 
PN is not detected towards the hot core, even though region "1" is partly overlapping with it. 
This allows us to rule out important formation pathways in hot gas.
\item The PN and SiO emissions are very similar in morphology and spectral shape, both having two strong emission peaks towards regions "1" and "2", while all sulphur bearing species emit predominantly from the hot core. 
\item We propose that the PN emitting region "2" is a "hot spot" for shocked material well separated from the hot core, that will allow us to study the chemistry of shocked gas without the influence of the hot core, similarly to other known chemically rich shocked regions powered by protostellar objects.
\item 
We derive excitation temperatures in the range $\sim 26 - 75$~K in region "1", and in the range $\sim 12 - 30$~K in region "2". The column density ratios of all species with respect to that of PN decrease by about an order of magnitude from region "1" to "2", except the SiO/PN ratio, which is constant within the uncertainties in both regions further indicating a common origin of the two species.
\item
We derive a (tentative) column density ratio PO/PN $\sim 1$ in region "2", in line with a pure shock model that does not need high cosmic ray ionisation rates.
\end{itemize}

An interesting follow up of our study will be to map transitions of PN at higher excitation to test whether these lines trace different (e.g. innermost?) material.
Moreover, observing more PN lines will allow us to derive \Tex\ for PN as well, and hence compute a more accurate estimate of $N_{\rm tot}$.

\begin{acknowledgements}
 We thank the anonymous  referee  for  their  careful  reading of  the  article  and  their  useful  comments.
C.M. acknowledges  funding from the European Research Council (ERC) under the European Union’s Horizon  2020  program,  through  the  ECOGAL  Synergy  grant  (grant  ID  855130).
 V.M.R. has received support from the project RYC2020-029387-I funded by MCIN/AEI /10.13039/501100011033, and from the the Consejo Superior de Investigaciones Cient{\'i}ficas (CSIC) and the Centro de Astrobiolog{\'i}a (CAB) through the project 20225AT015 (Proyectos intramurales especiales del CSIC).
I.J.-S. and L.C. acknowledge financial support through the Spanish grant PID2019-105552RB-C41 funded by MCIN/AEI/10.13039/501100011033.
I.J.-S., L.C. and V.M.R. acknowledge also financial support through the Spanish grant PID2022-136814NBI00 funded by MCIN/AEI/10.13039/501100011033 and by “ERDF A way of making Europe”.
S.V. ackowledges support from the European Research Council (ERC) Advanced grant MOPPEX 833460.
This  paper  makes  use  of  the  following ALMA data: ADS/JAO.ALMA$\#$2013.1.00489.S and ADS/JAO.ALMA$\#$2017.1.00501.S.ALMA is a partnership of ESO (representing its member states), NSF (USA)and NINS (Japan), together with NRC (Canada), MOST and ASIAA (Taiwan),and KASI (Republic of Korea), in cooperation with the Republic of Chile. TheJoint ALMA Observatory is operated by ESO, AUI/NRAO and NAOJ. 
\end{acknowledgements}

%
%

\end{document}